\documentclass[a4paper,11pt]{article}

\usepackage[utf8]{inputenc} % allows usage of spanish special characters
\usepackage[spanish,english]{babel} % english dictionary for proper hyphenation
\usepackage{amsmath,amsfonts,amssymb}% math expressions %changed Sudam
\usepackage{txfonts} % nice fonts 
\usepackage{authblk} % author affiliations
\usepackage{graphicx} % images
\usepackage{float} % image positioning
\usepackage{multirow} % allows merging cells in tables
\usepackage{makecell} % more table customization
\usepackage{arydshln}
\usepackage{placeins}
\usepackage{hyperref} % hyperlinks (remove if Nature-style, leave if Science-style)
\usepackage[
  backend=biber,
  style=numeric,
  sorting=none,
  sortcites=true,
  maxbibnames=99
]{biblatex}

\usepackage[table,xcdraw]{xcolor} % text and table background colors
\usepackage[margin=1.5cm]{geometry} % margins
\usepackage{lineno} % line numbers
\usepackage[misc]{ifsym} % use \Letter symbol for corresponding author
\usepackage{lipsum} % dummy text
\usepackage{sidecap} % figure captions on the side

\usepackage{comment}
\usepackage{booktabs}
\usepackage{multirow}
\usepackage{colortbl}
\usepackage{subcaption}
\usepackage[capitalize,noabbrev]{cleveref}

\usepackage{cancel}
\usepackage[symbols,nogroupskip,nonumberlist,nopostdot=false,abbreviations]{glossaries-extra}
\usepackage{array}
\usepackage{booktabs}
\usepackage{colortbl}
\usepackage{tikz} 
\usetikzlibrary{positioning, arrows.meta, calc}
\usepackage{soul}

\usepackage{tcolorbox}
\usepackage{siunitx}
\usepackage{pgfplots}

\usepackage{geometry}
\usepackage{algorithm}
\usepackage{algpseudocode}
\usepackage{siunitx}
\usepackage{enumerate}
\usepackage{enumitem}
\usepackage{caption}
\usepackage{tcolorbox}

\newcounter{myframecounter}
\newenvironment{myframe}[1]{%
    \refstepcounter{myframecounter}%
    \begin{tcolorbox}[colback=white,colframe=black,title=\textbf{Box \themyframecounter.} #1]\par\medskip
}{%
    \end{tcolorbox}
}
\crefname{myframecounter}{Box}{Boxes}
\newlist{myenumerate}{enumerate}{1}
\setlist[myenumerate]{label=\textbf{\arabic*.}}

\newlist{myitemize}{itemize}{1}
\setlist[myitemize]{label=\textbf{$\bullet$}}

\newlist{mydescription}{description}{1}
\setlist[mydescription]{font=\normalfont\bfseries}

\makeatletter\renewcommand{\@biblabel}[1]{#1.}\makeatother % change [number] for number. in references
\addto{\captionsenglish}{}
\hypersetup{ % FIXME: remove if Nature-style, leave if Science-style
  colorlinks = true,
  citecolor=  blue,
  linkcolor = {blue},
  filecolor = cyan, % controls color of external ref, if used
  urlcolor = blue
}
\DeclareCaptionLabelSeparator{vbar}{~\textbar~}
\DeclareCaptionLabelSeparator{dot}{.~}
\definecolor{lightgray}{gray}{0.9}
\newcommand{\figref}[2][]{%
  \hyperref[{#2}]{%
    Fig.~\ref*{#2}%
    \ifx\\#1\\%
    \else
      #1%
    \fi
  }%
}
\newcommand{\tableref}[1]{%
  \hyperref[{#1}]{%
   Table~\ref*{#1}%
  }%
}
\newcommand{\methodsref}[1]{%
  \hyperref[{methods:#1}]{%
   Methods:~\nameref*{methods:#1}%
  }%
}
\makeatletter % align title and affiliations to left
\renewcommand{\maketitle}{\bgroup\setlength{\parindent}{0pt}
\begin{flushleft}
  \textbf{\LARGE \@title}
  
  \bigskip
  
  \@author
\end{flushleft}\egroup
}
\makeatother
\renewenvironment{abstract} % align abstract to left
 {\par\noindent\textbf{\abstractname~\textbar}\ \ignorespaces}
 {\par\medskip}

\title{Learning virulence-transmission relationships using causal inference}

\author[1,2 \Letter]{Sudam Surasinghe}
\author[1,2,3 \Letter]{C. Brandon Ogbunugafor}
\affil[1]{Department of Ecology \& Evolutionary Biology,
Yale University, New Haven, CT 06520, USA}
\affil[2]{Public Health Modeling Unit, Yale School of Public Health, New Haven, CT 06510, USA}
\affil[3]{Santa Fe Institute, Santa Fe, NM 87501, USA}

\affil[$\textrm{\Letter}$]{S.S.: sudam.surasinghe@yale.edu; C.B.O: brandon.ogbunu@yale.edu}
\date{}
  
\begin{document}
%\linenumbers
\begin{refsection}
\maketitle

\bigskip\bigskip

\begin{abstract}

The relationship between traits that influence pathogen virulence and transmission is part of the central canon of the evolution and ecology of infectious disease. However, identifying directional and mechanistic relationships among traits remains a key challenge in various subfields of biology, as models often assume static, fixed links between characteristics. Here, we introduce \emph{learning evolutionary trait relationships} (LETR), a data-driven framework that applies Granger-causality principles to determine which traits drive others and how these relationships change over time. LETR integrates causal discovery with generative mapping and transfer-operator analysis to link short-term predictability with long-term trait distributions. Using a synthetic myxomatosis virus-host data set, we show that LETR reliably recovers known directional influences, such as virulence driving transmission. Applying the framework to global pandemic (SARS-CoV-2) data, we find that past virulence improves future transmission prediction, while the reverse effect is weak. Invariant-density estimates reveal a long-term trend toward low virulence and transmission, with bimodality in virulence suggesting ecological influences or host heterogeneity. In summary, this study provides a blueprint for learning the relationship between how harmful a pathogen is and how well it spreads, which is highly idiosyncratic and context-dependent. This finding undermines simplistic models and encourages the development of new theory for the constraints underlying pathogen evolution. Further, by uniting causal inference with dynamical modeling, the LETR framework offers a general approach for uncovering mechanistic trait linkages in complex biological systems of various kinds.

\end{abstract}

\section{Introduction}
Since the early 1980s, the virulence transmission tradeoff hypothesis has provided a foundational framework in the evolution and ecology of infectious diseases. It proposes that virulence (``the harm a pathogen does to its host'') \cite{longdon2015causes} evolves in relation to transmission up to an optimal threshold, beyond which elevated host mortality undermines transmission, and consequently, pathogen fitness \cite{anderson1982coevolution, ewald1983host, lenski1988evolution, bull1994virulence, read1994evolution}. Despite its conceptual and pedagogical appeal, empirical studies reveal that this relationship is deeply \emph{context-dependent}: correlations observed under one set of host-parasite conditions often do not hold under others \cite{alizon2009virulence, kabengele2024meta, rafaluk-mohr_relationship_2019, acevedo2019virulence}. This dubiousness has encouraged the field to reflect on the assumptions underlying the relationship between virulence and transmission. That is, epidemic behaviors perceived as the product of evolutionary forces could instead be ecological or behavioral in nature \cite{fussmann2007eco, bharti2021linking}. The idiosyncratic nature of the relationship between virulence and transmission underscores the greater need for methods that can better capture the mechanistic relationship between biological traits. Specifically, the limitations of univariate tradeoff analysis highlight the need for multivariate causal methodologies that transcend the study of disease, and can be applied across many subfields of evolutionary biology and ecology. 
  
Causal drivers of quantitative traits are central yet often ambiguously treated in genetics, particularly for complex traits \cite{page2003we,kurth2021continuing}. Existing approaches mostly address static causality within Pearl’s \cite{pearl2010causal} framework, while explicit causal studies remain scarce \cite{mackay2009genetics,battle2014determining,reitan2019layeranalyzer}. Time-series perspectives inspired by Granger’s predictive causality \cite{granger1969investigating,Wiener1956} offer dynamic alternatives but are rarely applied to evolutionary scenarios \cite{guo2010granger,deyle2016global}. In epidemiology, causal modeling has advanced toward the integration of environmental, social, and biological determinants \cite{eisenberg2007environmental,susser1996choosing,koopman1999individual,saul2017causal,krieger2001theories}. Studies often adopt Rubin’s potential outcomes framework to estimate intervention effects \cite{rubin1972estimating,halloran2012causal,vanderweele2012components}, while extensions such as marginal structural models address time-varying exposures \cite{eisenberg2003bias,robins2000marginal,halloran1995causal}. Recognizing that environmental drivers vary through time, time-series causal discovery is particularly valuable in the study of ecological systems. Methods such as convergent cross-mapping and related data-driven causal tools have been used to detect driver response relationships in ecosystem time series \cite{sugihara2012detecting,liu2022environmental}.
In parallel, Granger-style analyses have been applied to infer environmental and behavioral drivers of incidence \cite{prabodanie2020coherence,hossain2024analysis,grande2023granger,sebayang2025unraveling,romero2023understanding}, highlighting the complementarity between predictive and interventionist causal frameworks. 

Despite these advances, there remains a methodological gap in identifying causal relationships between dynamic biological traits. Such tools are essential for clarifying the elusive connection between virulence and transmission and, more broadly, for uncovering causal structure in biological systems. To address this, we introduce the \emph{learning evolutionary trait relationships} (LETR) framework, which combines predictive causality in the Granger tradition with discrete dynamical systems. LETR employs modern information-theoretic extensions—such as transfer entropy, causation entropy, and geometric information flow \cite{schreiber2000measuring,sun2014causation,surasinghe2020geometry}—to capture nonlinear and context-dependent interactions among traits associated with virulence and transmission in this setting.  Evolutionary updates are represented as discrete generational maps, which naturally accommodate fixed points, bifurcations, and chaotic behavior \cite{may1976simple,koella1999population}. Using mechanistic examples, such as the myxomatosis model of Dwyer et al. \cite{dwyer1990simulation}, LETR can illustrate how virulence can act as a causal driver of transmission through within-host effects on parasite load.

We offer several original contributions. Methodologically, we present a scalable, data-driven pipeline that identifies Granger-style directional drivers from multivariate time series, fits generative update maps conditioned on those drivers, and analyzes the corresponding transfer operator to recover long-term trait distributions. The LETR framework therefore enables tests of which measured traits act as upstream causal drivers and yields population-level distributions that link mechanistic hypotheses about traits and ecological processes to empirically observable patterns. With regard to the canon in the evolution and ecology of infectious diseases, we demonstrate that static tradeoff models are insufficient for properly deconstructing the relationships between pathogen traits. Alternatively, the field should consider new theoretical models that incorporate the dynamism and context-dependence of trait relationships governing pathogen evolution.

 By introducing LETR, we bridge the gap between empirical time series and theoretical models of trait correlation, providing a quantitative language for disentangling directionality and feedback in evolutionary and ecological dynamics. In doing so, we advance a broader agenda that transcends disease and applies across evolutionary biology and ecology: to move beyond static models of trait correlation, and toward a dynamic understanding of how causal linkages among traits determine long-term ecological equilibria, evolutionary trajectories, and the predictability of complex biological systems.

\section{Methods}

\begin{figure}
    \centering
    \includegraphics[width=1\linewidth]{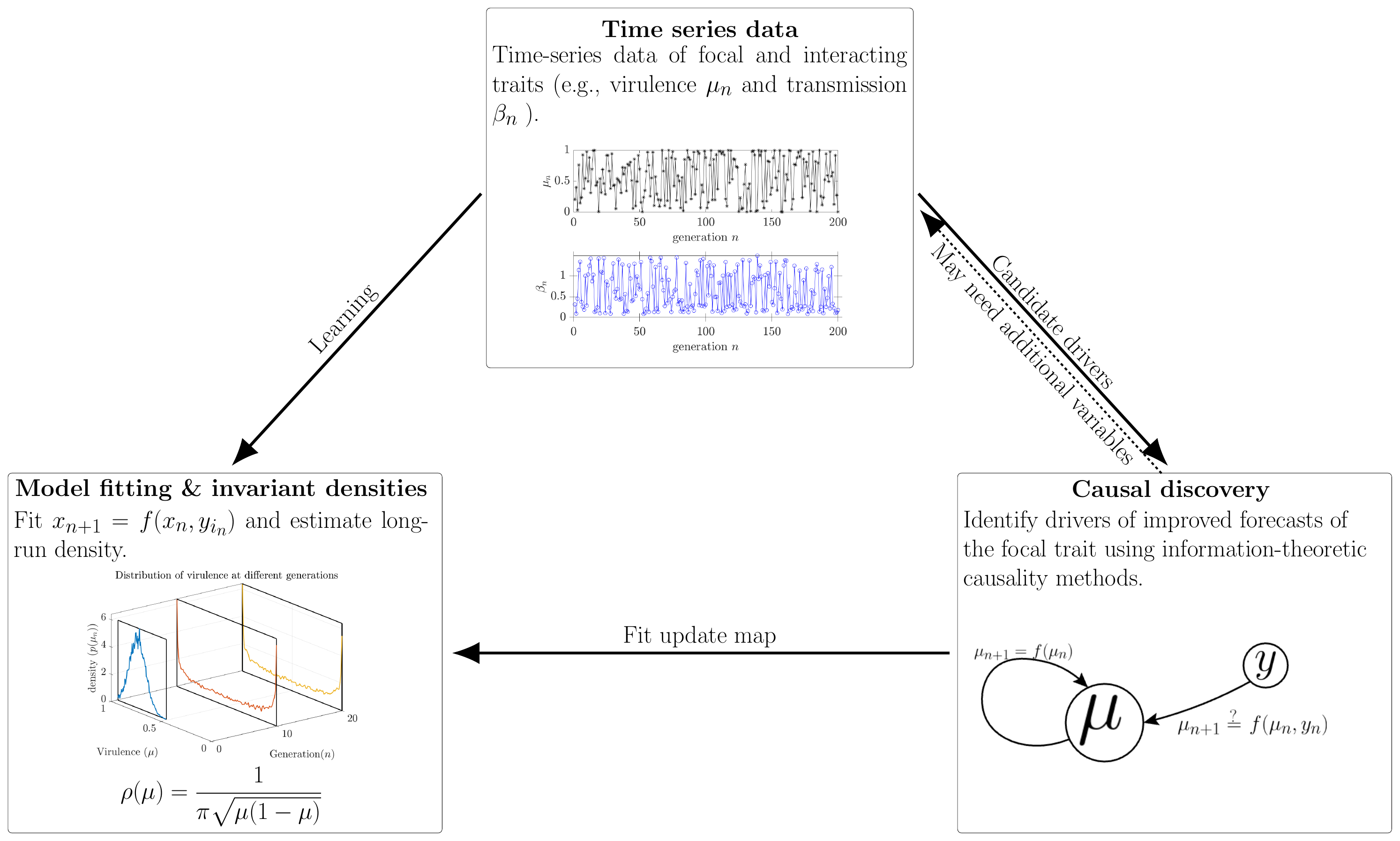}
    \caption{%
\textbf{Illustration of the \textit{learning evolutionary trait  relationships} (LETR) framework.} 
The top box shows time series of focal and interacting traits, such as virulence ($\mu_n$) and transmission ($\beta_n$), collected from evolving pathogen populations or from host-level surveillance. 
\textbf{Causal discovery} identifies which measured variables improve short-term forecasts of a chosen focal trait by testing whether adding a candidate predictor, such as an environmental covariate, a demographic or behavioral variable, a genetic marker, or a treatment exposure improves one-step forecasts. Methods include geometric information flow (GeoC) \cite{surasinghe2020geometry}, transfer entropy \cite{schreiber2000measuring} and causation entropy \cite{sun2014causation}. 
\textbf{Model fitting} fits a generative update map $x_{n+1}=f(x_n,y_{i_n})$ conditioned on drivers discovered in step one, where $f$ may be a pedagogical map such as the logistic map or a regularized supervised learner chosen by cross-validation, and where fitted models yield interpretable summaries and partial dependence visualizations for applied researchers. 
\textbf{Invariant densities} analyze ensemble behavior under $f$ by approximating the Frobenius-Perron transfer operator and estimating the invariant density of $x$, which summarizes the long-run distribution of the focal trait under the inferred dynamics and observed driver processes, highlights prevalent phenotypes and multimodality, and indicates how interventions may reshape stable outcomes. 
Arrows indicate the workflow and feedback, with dashed links from causal discovery back to the data denoting the possibility of adding additional variables to improve inference, and arrows from data to model fitting denoting direct learning of long-term behaviour from observed trajectories.}
    \label{fig:Process}
\end{figure}

In this study, we propose a data-driven toolkit for learning trait dynamics from longitudinal observations. Here, ``learning'' refers to the process of inferring underlying patterns and causal relationships from data, as commonly used in machine learning and data-driven analysis, rather than traditional cognitive conceptions of learning. Modern monitoring and aggregation efforts are steadily increasing the temporal resolution and trait coverage available to epidemiologists and disease ecologists, enabling inference that moves beyond simple summary statistics and fixed distributional assumptions. The \emph{learning evolutionary trait relationships} (LETR) framework exploits these richer time series to identify which measured traits functionally drive others and to condition on discovered drivers when fitting generative update maps. In practice, LETR is a two-stage, data-driven pipeline (see \cref{fig:Process}). First, we identify Granger-predictive drivers by testing whether candidate variables improve one-step forecasts of a focal trait using geometric information flow (GeoC) \cite{surasinghe2020geometry} or alternative Granger-style measures such as transfer entropy \cite{schreiber2000measuring} or causation entropy \cite{sun2014causation}. Second, we fit a conditional generative update map of the form 
\(
\mu_{n+1} = f(\mu_n, y_{i_n}),
\)
using interpretable learners or regularized supervised methods, and then analyze ensemble evolution under \( f \) by approximating the Frobenius-Perron transfer operator to obtain invariant densities and long-term trait distributions \cite{lasota1998chaos}.

\subsection{Identify Granger causal variables}
We are interested in identifying the causal variables associated with the trait of interest. In this study, we focus on virulence, denoted as $\mu$, as the trait of interest. However, readers may substitute virulence with any trait relevant to their study.

The virulence data may not be fully explained by its previous generation value. Therefore, we are interested in understanding the potential influence of other Granger causal variables on virulence. Granger causality explores whether one time series can predict the behavior of another time series (see Supporting Information Figure S1). In other words, we are interested in exploring the following question: ``Is an additional variable \(y\) required to elucidate the relationship between the present and past data of a variable \(x\) (in this case, virulence)?"

This question involves distinguishing between two alternative versions of the potential origins of \(\mu_{n+1}\) and is based on choosing one of the following two cases:
\begin{equation}\label{decide}
    \mu_{n+1}= f(\mu_n),\ \ \mbox{ or }\ \ \mu_{n+1}= f(\mu_n,y_n),
\end{equation}
to describe the actual function \(f\). We use the shorthand notation: 
\begin{equation}\label{CauQs}
\mu_{n+1} =\hspace{-0.08in} ? \hspace{0.08in} f(\mu_n, y_n)
\end{equation} 
to represent the above statement. We also employ a geometric measure for causality, \(GeoC(.)\) \cite{surasinghe2020geometry}, to determine the conditional information flow, as measured by the difference in the correlation dimension of the data sets [\(\mu_{n},\mu_{n+1}\)] and \(\mu_{n}\):

\begin{equation}\label{ConGeoC}
GeoC(\mu_{n+1}|\mu_n)=D_2([\mu_{n},\mu_{n+1}]) -D_2(\mu_{n}),
\end{equation}  
where \(D_2(.)\) represents the correlation dimension of the given set. The geometric measure for causal inference from $y$ to $\mu$ can therefore be computed as 
\begin{equation}\label{GeoCytomu} 
GeoC_{y\to \mu} = GeoC(\mu_{n+1}|\mu_n) - GeoC(\mu_{n+1}|\mu_n, y). 
\end{equation}

If virulence is not influenced by any additional causal variable, then $GeoC(\mu_{n+1}|\mu_n)$ should be approximately zero, or both $\mu_n$ and $[\mu_n, \mu_{n+1}]$ reside on manifolds of the same dimension. In this case, adding any additional variable $y$ does not change the value of the conditional geometric information, i.e., $GeoC(\mu_{n+1}|\mu_n, y)$ remains the same as $GeoC(\mu_{n+1}|\mu_n)$. Hence, $GeoC_{y \to \mu} \approx 0$ for any $y$ in this scenario.

When $GeoC(\mu_{n+1}|\mu_n) \gg 0$, it implies the existence of at least one causal variable and highlights the necessity to update the model by identifying such a variable. In such instances, expert reasoning or prior knowledge can be employed to consider potential causal variables, which can then be added to the existing dataset by incorporating their respective time series data. Subsequently, the conditional geometric information flow of the new dataset can be analyzed to assess the effect of the considered causal variable. It is worth noting that multiple causal variables may influence virulence, and a systematic evaluation of geometric information flow could help identify them. \cref{fram:Vir} summarizes this process as a framework.

\begin{myframe}{Framework for learning virulence (the \textit{trait of interest}) }\label{fram:Vir}
    \begin{myenumerate}
        \item \textbf{Data:} Utilize time series data of $N \times 1$ vector $\mu$  to construct vectors $\mu_n=\mu(1:end-1)$ and $\mu_{n+1}=\mu(2:end)$.
        \item Compute the correlation dimension (\(D_2(\cdot)\)) and assess the conditional geometric information flow: \(GeoC(\mu_{n+1}|\mu_n) = D_2([\mu_{n},\mu_{n+1}]) - D_2([\mu_{n}])\).
        \item \textbf{If}   \(GeoC(\mu_{n+1}|\mu_n)\approx 0\): %or $\left\lceil D_2([\mu_{n}\ \mu_{n+1}])\right\rceil =\left\lceil D_2([\mu_{n}])\right\rceil $ (where $\left\lceil .\right\rceil$ is the ceiling function):
        \begin{myitemize}
        \item There are no effective causal variables for $\mu$; hence, $\mu_{n+1}$ can only be modeled by $\mu_n$.
        \item Identify a discrete map \(f\) that is fitted to the data, satisfying the relationship \(\mu_{n+1} = f(\mu_n)\). For instance, consider the logistic map \(f_a(\mu)=a\mu(1-\mu)\) with a parameter \(a\).
    \item Leveraging insights into the map's qualitative behavior, interpret how virulence evolves over generations. Explore whether it tends to settle into a stable fixed point, follow a periodic pattern, exhibit a chaotic trajectory, or display other dynamic behaviors. 
    \item Utilize transfer operators to analyze the temporal changes of population densities related to virulence.
    \end{myitemize}
   
    \item \textbf{else} 
\begin{myitemize}
        \item Consider a causal variable \( y \) for \( \mu \) through expert reasoning or prior knowledge.
\item Then, incorporate time series data \( y_n \) into the current dataset of \( \mu_n \).
\item Now, utilize the datasets \( [\mu_{n+1}|\  \mu_n,\ y_n] \) and \( [\mu_n,\  y_n] \) to assess the conditional geometric information flow \(GeoC(\mu_{n+1}|\mu_n, y_n) = D_2([ \mu_{n+1},\ \mu_{n},\ y_n]) - D_2([\mu_{n},\ y_n])\).
\item \textbf{if} \(GeoC(\mu_{n+1}|\mu_n,\ y_n)\approx 0\): %or $\left\lceil D_2([\mu_{n}\ \mu_{n+1}\ y_n])\right\rceil =\left\lceil D_2([\mu_{n}\ y_n])\right\rceil $
\begin{itemize}
    \item Only causal variable is $y$ and use supervised learning methods to fit the data to the model $\mu_{n+1}=f(\mu_n,y_n)$.  
    \item Utilize transfer operators to analyze the temporal changes of population densities related to virulence.
\end{itemize}
\item \textbf{else}
 \begin{itemize}
     \item Consider another causal variable \( z \) and repeat the process until all the causal variables have been identified.
 \end{itemize}
\end{myitemize}
     \end{myenumerate}
\end{myframe}

\subsection{Learning virulence using discrete maps and transfer operators}
 Virulence is a canonical idea within various subfields of infectious diseases, often defined as the harm a pathogen inflicts on its host. It can be measured in different ways across host-parasite systems, with implications for how we identify relationships between it and other traits \cite{kabengele2024meta, acevedo2019virulence}.  In this study, we use the host population's mortality rate attributable to the pathogen's influence (often summarized in public health as case fatality rate, CFR) as a proxy for virulence. While this definition might be very specific, the virulence literature has been famously dubious about how to measure it in host-parasite systems \cite{kabengele2024meta}. Consequently, to minimize ambiguity in how we define virulence, we define host death as the ultimate form of harm to a host. We hypothesize that virulence undergoes temporal variations driven by underlying changes in the pathogen's characteristics. For demonstration purposes, we utilize a simplified model of virulence that excludes additional causal variables. A numerical illustration using the logistic map, together with poliovirus mortality data, is provided in Supporting Information Section S2. The virulence of the pathogen at its $n^{\text{th}}$ generation is denoted by $\mu_n$, and is constrained to the unit interval. To model the dynamics of virulence, deterministic or random maps can be employed. Here, our focus lies on deterministic maps, specifically represented by the map $f$ operating on the unit interval $[0, 1]$:
\begin{eqnarray}\label{map}
f:[0,1]  &\rightarrow& [0,1] , \nonumber \\
 \mu &\mapsto& f(\mu).
\end{eqnarray}
The qualitative behavior of these maps can be employed to model the dynamics of virulence. Evolutionary and ecological dynamics embodied in host-parasite interactions can give rise to various dynamical behaviors of virulence, such as asymptotically stable fixed points, periodic orbits, or even chaotic trajectories. 

When the map $f$ is chaotic, analysis often involves considering an ensemble of many initial conditions, rather than following a forward time orbit \(\mathbb{O^+}=\{\mu,f(\mu),f^2(\mu),...\}\) from a single initial condition \cite{bollt2000controlling}. This scenario can be analyzed using a transfer operator known as the Frobenius-Perron operator. This operator is employed to describe the orbit of ensembles when they are distributed in $p\in L^1([0,1])$, and the map $f$ is a nonsingular transformation and measurable relative to a Borel sigma algebra set. The Frobenius-Perron operator is a linear map, ${\cal P}_f:L^1([0,1])\rightarrow L^1([0,1])$, which follows the discrete continuity equation, 
\begin{equation}
    \int_B p_{n+1} d\nu=\int_{f^{-1}(B)} p_n d\nu, \mbox{ for any } B\in \mathbb{B}, \label{FPEq}
\end{equation}
where $\mathbb{B}\subseteq [0,1]$ is the Borel sigma algebra of measurable sets with measure $\nu$, and  $p_n$ denotes the $n^{\text{th}}$ iteration of the initial density. For differentiable maps, this simplifies to:
\begin{equation}
    {\cal P}_f[p](\mu)=\sum_{y:\mu=f(y)} \frac{p(y)}{|Df(y)|},
\end{equation}
where, if \(f(y)\) is a univariate function, \(|Df(y)|=|f'(y)|\) represents the absolute value of the derivative. For multivariate functions, it corresponds to the determinant of the Jacobian matrix. Additionally, the invariant density (fixed point) of the operator, expressed as \(\pi\):
\begin{equation}
    {\cal P}_f[\pi]=\pi\label{FPEqIn},
\end{equation}
is often of interest for understanding the long-term behavior of the variable of interest \cite{bollt2000controlling}. Here, we use the transfer operator to elucidate the behavior of virulence (variable of interest) when the logistic map undergoes chaotic behavior. 

\section{Results}
In this section, we present the analyses of two case studies that provide methodological validation and empirical application for the LETR framework. Specifically, we structure the results as follows:

\begin{itemize}
  \item \textbf{Case study I:} Method validation using synthetic data from a myxomatosis model, in which we test whether LETR recovers known directional influences inferred from the model (\cref{sec:CausalModel}).
  \item \textbf{Case study II:} Applied analysis of worldwide epidemiological data on SARS-CoV-2, in which we perform Granger style causal discovery, fit conditional generative maps, and estimate invariant densities for virulence and transmission (\cref{sec:CovidMain}).
\end{itemize}

Here, we assume that only a single strain of the pathogen enters a given host, and that the pathogen exits the host as a single strain, which may differ from the entering strain. In our analyses, we assume the transmission rate, \(\beta_n\), and virulence, \(\mu_n\), to be dynamic variables. Subscripts denote the generation (or time step); for instance, \(\beta_n\) denotes the transmission rate of the pathogen in the \(n^\text{th}\) generation (or time step). We also refer readers to Supporting Information Figure S6 for a graphical illustration of these assumptions. Figure S6 illustrates the process by which a pathogen in the \(n^\text{th}\) generation, characterized by parameters \(\mu_n\) and \(\beta_n\), enters a host and undergoes mutations and ecological changes. As a result, the pathogen exits the host with updated parameters \(\mu_{n+1}\) and \(\beta_{n+1}\) in the \((n+1)^\text{th}\) generation. However, when the data lacks a clearly defined generational structure, we treat each unit of time as a single generation and denote $n$ as the corresponding time step.

The goal is to model the dynamics of these variables, assuming that the $(n+1)^{\text{th}}$ generation is influenced solely by the $n^{\text{th}}$ generation values. This assumption implies that mutations and ecological factors within the current host determine the characteristics of the pathogen as it exits the host. This process was modeled using the concept of a discrete map. In classical modeling approaches, descriptive statistics are used to represent parameter values or to trace the path from a single initial condition. In this study, we examine the dynamics of ensembles of initial values for the parameters of pathogens under a discrete map that represents evolutionary and ecological changes (see supporting information Figure S7). In the results section, we present a detailed analysis of causal discovery using synthetic benchmarks (Case 1) and worldwide epidemiological data on a modern pandemic of SARS-CoV-2 (Case 2). We also provide data-driven estimates of the invariant densities of virulence and transmission for the global SARS-CoV-2 record.

\subsection{Case study I: Causal model for virulence and transmission in a synthetic dataset of myxomatosis}
\label{sec:CausalModel}
The story of myxomatosis provided scientists with an unprecedented opportunity to observe and document the real-time co-evolution of both host and pathogen in natural populations \cite{fenner1965myxomatosis, kerr2012myxomatosis}. When the highly virulent myxoma virus was released into Australian rabbit populations in 1950 and European rabbit populations in 1952, it initially caused extremely high mortality rates, but less virulent strains evolved, while rabbits simultaneously evolved genetic resistance \cite{fenner1953changes, marshall1958studies, fenner1965myxomatosis}. This system produced one of the greatest examples of rapid evolutionary change ever recorded and demonstrated that virulence evolution involves complex tradeoffs between transmissibility and host survival \cite{kerr2012evolutionary}. In light of this, we utilize it as a case study for our exploration of the utility of the LETR method.

\paragraph{Synthetic data motivated by myxomatosis.}
Dwyer et al.\ \cite{dwyer1990simulation} provide the mechanistic motivation for our synthetic experiments. In their myxomatosis model the within-host virus concentration follows a Ricker form \(v(t)=k_1 t e^{-k_2 t}\) and the fitted parameters \(k_1,k_2\) are regressed on strain case mortality so that virulence maps to the timing and magnitude of within-host infectiousness \cite{dwyer1990simulation}. Empirically, the probability that a vector acquires infection from an infected host increases with within-host viral titer.  Consequently, changes in case mortality propagate through the chain \(\mathrm{CM}\mapsto (k_1,k_2)\mapsto v(t)\mapsto\) transmission probability \cite{dwyer1990simulation,fenner1956epidemiological}. Dwyer and colleagues showed that modest shifts in host life history, seasonal forcing or resistance can move the system among equilibria, sustained cycles and period doubling, which demonstrates that virulence time series may be complex even when the underlying rules are deterministic and that past virulence can therefore carry information about future transmission. These mechanistic links motivated our synthetic design, which emulates non-Gaussian trait densities, a strongly self predictive virulence trajectory, and an asymmetric causal structure in which past virulence improves prediction of future transmission while the reverse effect is weak. To construct a controlled benchmark we simulated:
\begin{equation}
\begin{aligned}
\mu_{n+1} &= 4\,\mu_n(1-\mu_n),\\
\beta_{n+1} &= 1.5\,\operatorname{sech}^2\!\bigl(\mu_n+\beta_n\bigr).
\end{aligned}
\label{eq:synthetic_maps}
\end{equation}
The resulting time series are summarized in supporting information Figure S8 and in supporting information Section S2.2.3. 

\subsubsection{Simulated virulence and transmission dynamics}

\begin{figure}[ht]
    \centering
         \begin{subfigure}[b]{0.45\textwidth}
             \centering
    \includegraphics[width=\textwidth]{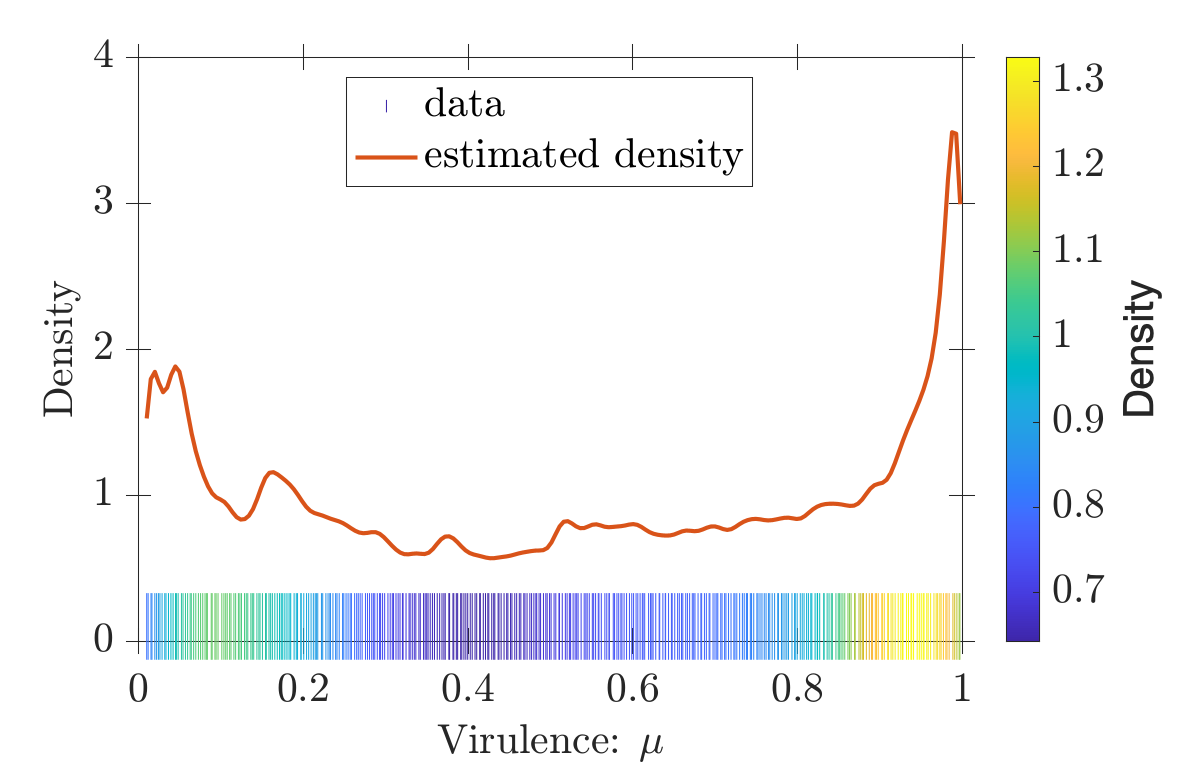}
    \caption{}
    \label{fig:DenMu}
         \end{subfigure}
     %\hfill
    \begin{subfigure}[b]{0.45\textwidth}
         \centering
\includegraphics[width=\textwidth]{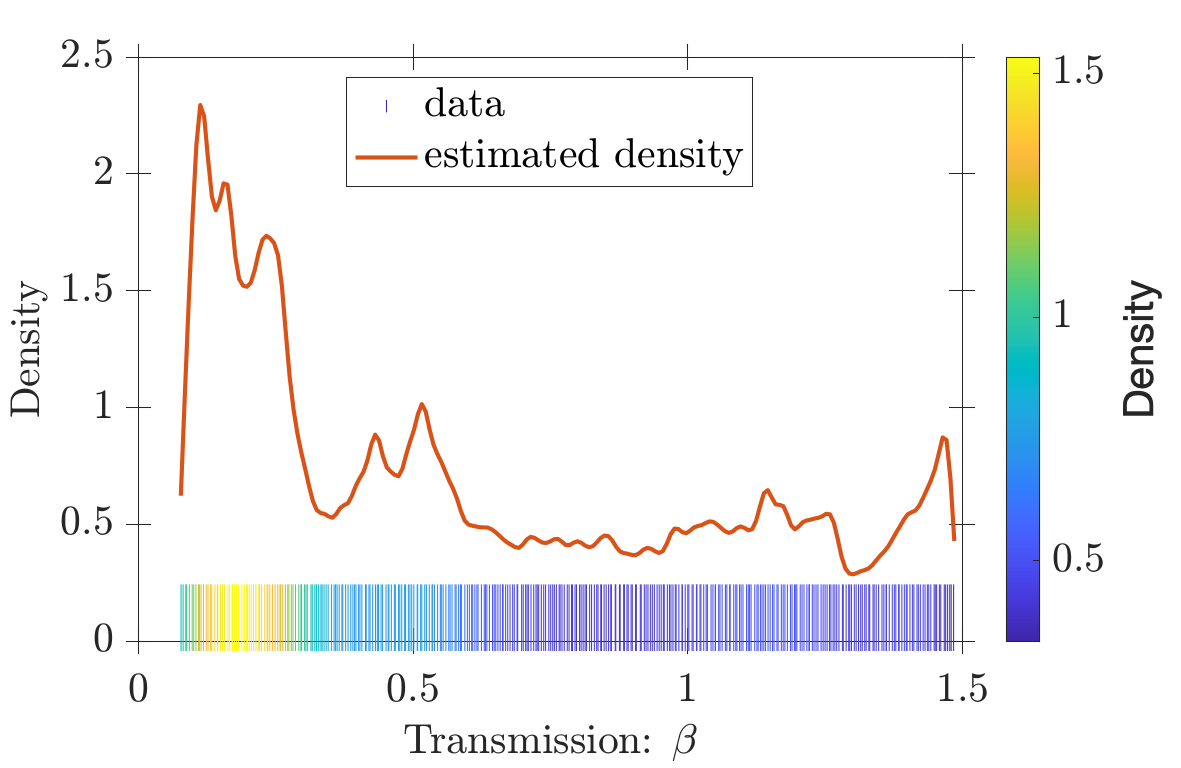}
    \caption{}
    \label{fig:DenBeta}
     \end{subfigure}
     
         \begin{subfigure}[b]{0.45\textwidth}
         \centering
\includegraphics[width=\textwidth]{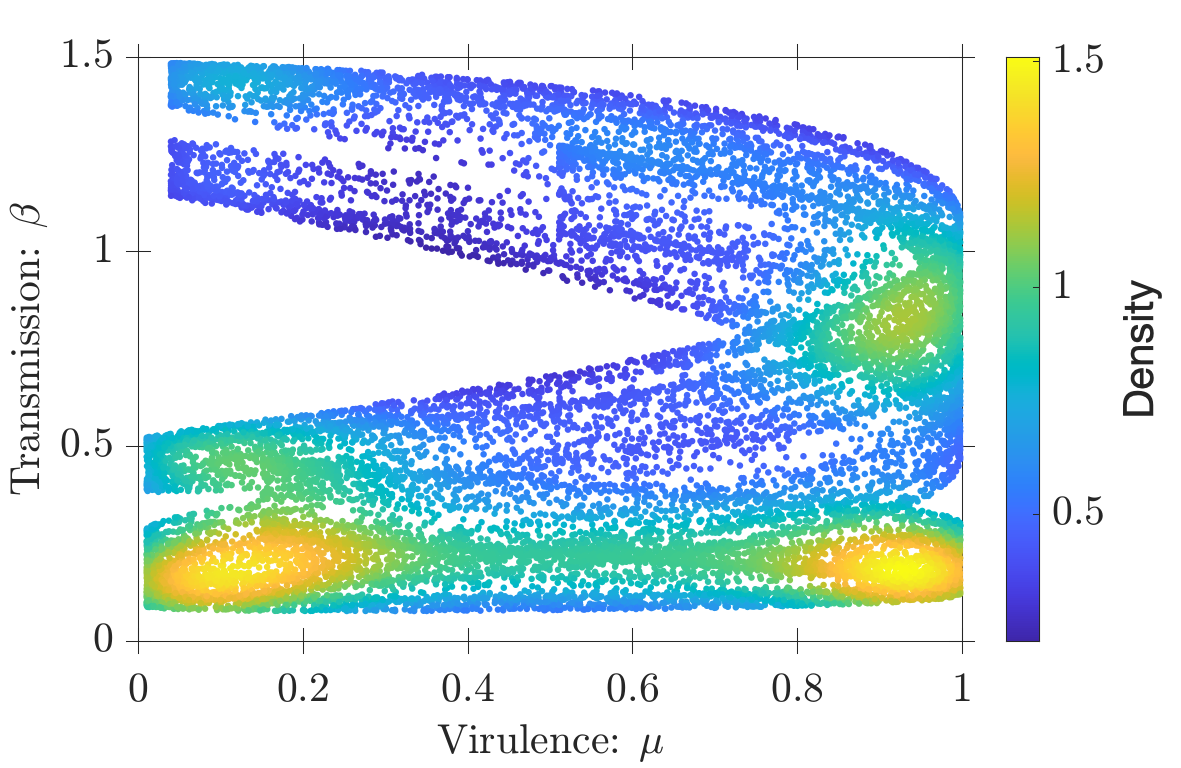}
    \caption{}
    \label{fig:DenMuBeta}
     \end{subfigure}
\caption{ \textbf{Simulated complex relationship between virulence and transmission, motivated by myxomatosis.} We illustrate the complex relationship between virulence and transmission trait dynamics using a methodological benchmark. (a) The marginal density of virulence \(\mu\) estimated by kernel density estimation and departs markedly from a Gaussian form, exhibiting skewness and heavy tails that indicate a nontrivial probability of extreme virulence values. (b) The marginal density of transmission \(\beta\) is likewise skewed and not well summarized by mean and variance alone. (c) The joint density of \(\mu\) and \(\beta\)reveals a nonlinear, non-monotonic, and complex relationship between the traits. The dataset was constructed to avoid Gaussian assumptions and a simple tradeoff structure, providing a demanding test case for causal discovery methods. The complex dynamics underscore the need for flexible, data-driven causal inference when interpreting biological trait relationships.}
         \label{fig:DensitySolo}
\end{figure}

 \cref{fig:DensitySolo} highlights key aspects of the diversity observed in both virulence and transmission rate within the simulated model. Using density plots, we present the distributions of these two critical parameters, emphasizing their variability and departures from conventional assumptions, such as normality. As illustrated in panel (c), the relationship between virulence and transmission rate deviates from the simplistic correlational patterns often assumed in many standard models from ecology and evolutionary biology (e.g., life history theory, the evolution of virulence) \cite{stearns1998evolution, bull1994virulence}. This deviation underscores the need to employ more advanced analytical methods that extend beyond basic tradeoffs or associations. Consequently, it serves as a demonstrative example of how our framework can be applied to real-world experimental or observational data, offering deeper insights into invariant densities and the causal mechanisms driving dynamic processes. 
\subsubsection{Visualizing simulated data to identify Granger causal relationships}\label{sec:DVvbeta}
\begin{figure}[ht!]
    \centering
         \begin{subfigure}[b]{0.45\textwidth}
             \centering
    \includegraphics[width=\textwidth]{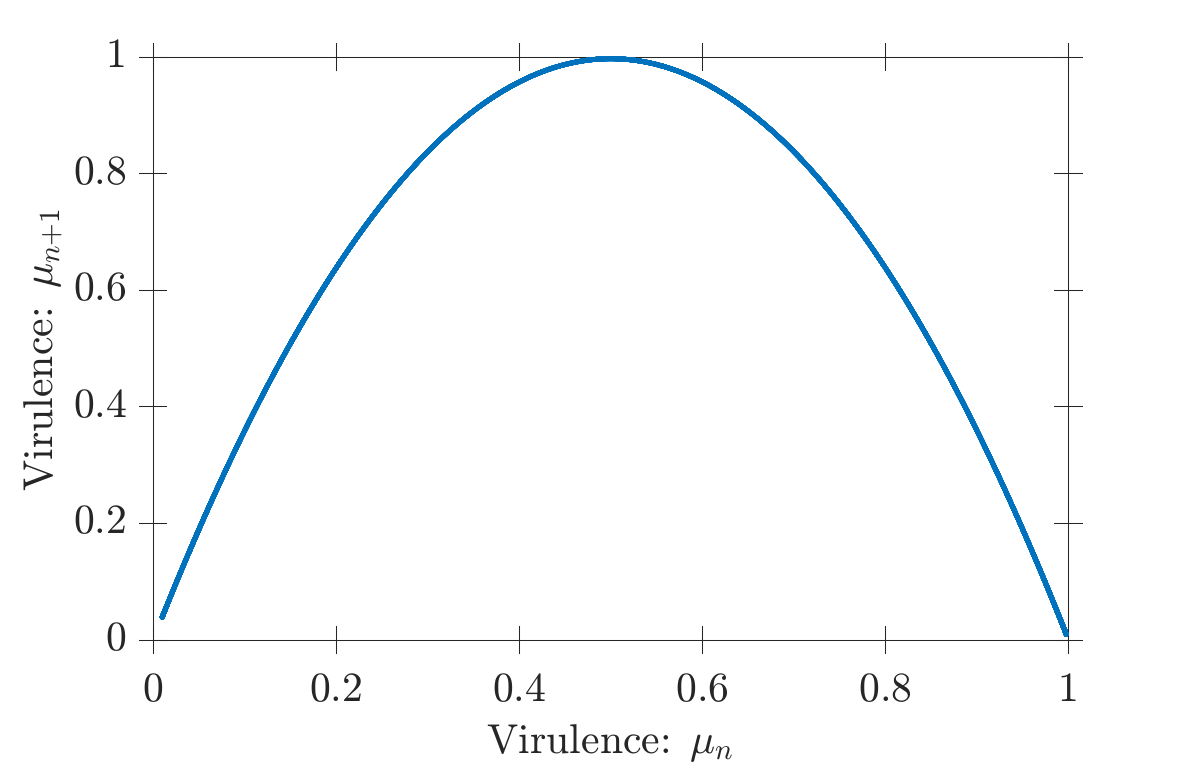}
    \caption{}
    \label{fig:muTser}
         \end{subfigure}
   
    \begin{subfigure}[b]{0.45\textwidth}
         \centering
\includegraphics[width=\textwidth]{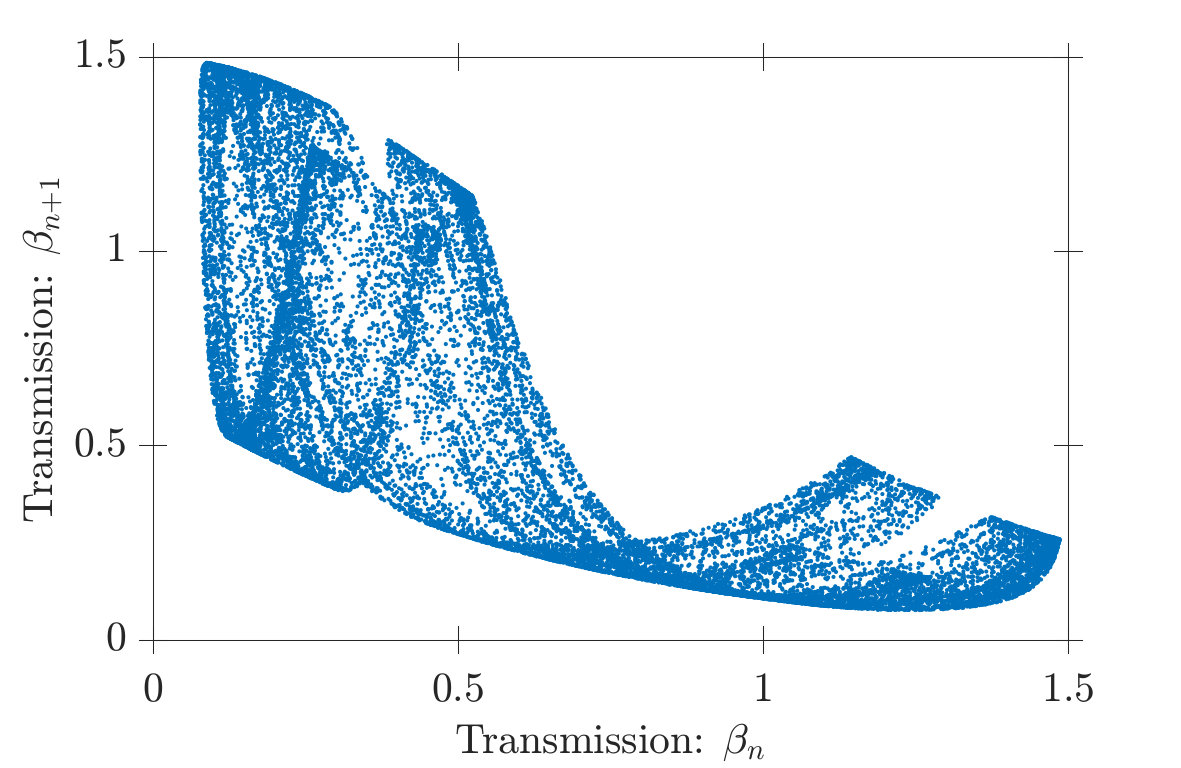}
    \caption{}
    \label{fig:btSer}
     \end{subfigure}
     %\hfill
         \begin{subfigure}[b]{0.45\textwidth}
         \centering
\includegraphics[width=\textwidth]{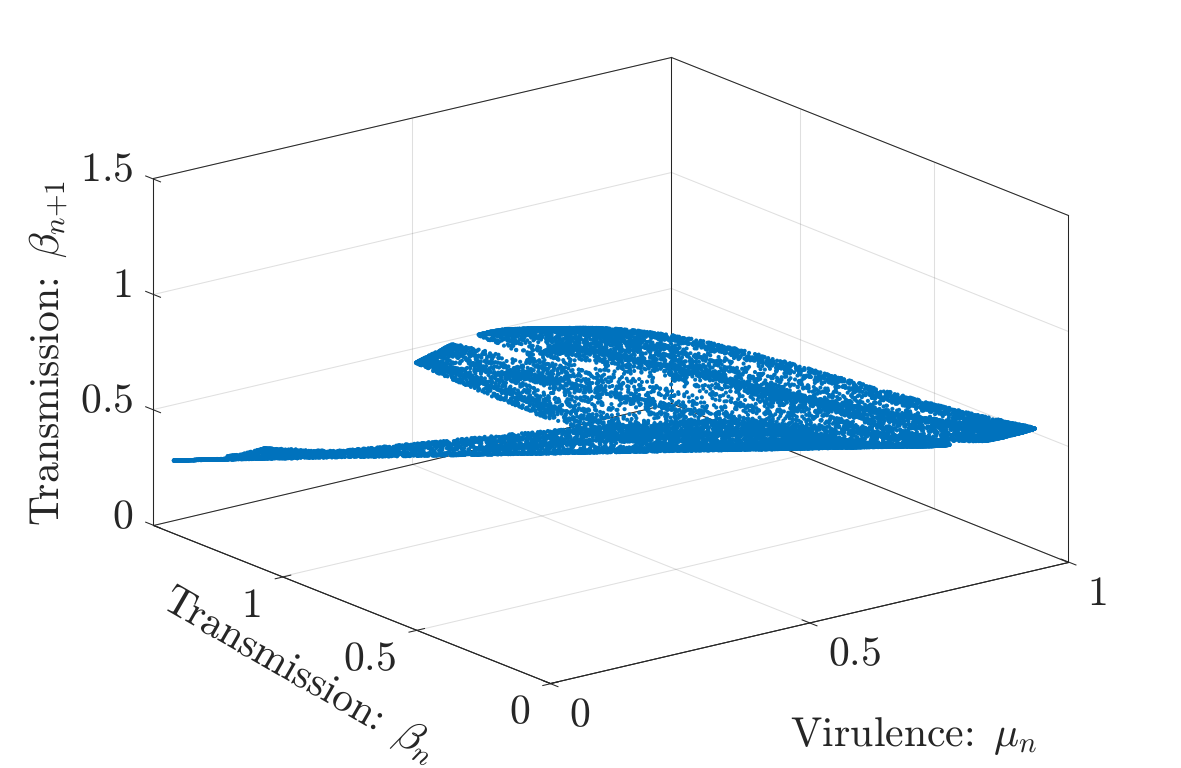}
    \caption{}
    \label{fig:munBetanBetaP}
     \end{subfigure}
        \caption{\textbf{ Geometric diagnostics of the virulence–transmission causal model based on synthetic data inspired by myxomatosis.}
One-step relationships between consecutive generations illustrate that past virulence improves the prediction of future transmission. 
(a) Scatter plot of \(\mu_n\) versus \(\mu_{n+1}\), showing a tight low-dimensional relationship consistent with strong self-predictability of virulence and the existence of a parsimonious update map for \(\mu\). 
(b) Scatter plot of \(\beta_n\) versus \(\beta_{n+1}\), showing weak one-step predictability when using \(\beta\) alone. The point cloud lies in a higher-dimensional space than a simple one-dimensional curve. The estimated correlation dimension for the pair \([\beta_n,\beta_{n+1}]\) is approximately 1.17, indicating that \(\beta\) dynamics are not adequately described by a one-dimensional map in \(\beta\) and that additional drivers are required. 
(c) Scatter plot of \((\mu_n,\beta_n)\) versus \(\beta_{n+1}\), showing that the point cloud lies on an effectively two-dimensional manifold, thereby recovering the structure absent from panel (b). 
Together, the contrast between panels (a) and (b) and the structural recovery in panel (c) provide visual and model-free evidence that \(\mu_n\) carries predictive information for subsequent transmission. Quantitative confirmation of these observations is provided by geometric causal inference and correlation dimension estimates presented in the main text. This figure illustrates the geometric validation of the recovery of the hidden benchmark causal structure and captures the expected virulence–transmission dynamics with complex, biologically relevant trait interactions.}
 
         \label{fig:tSerBoth}
\end{figure}

Following the framework suggested in \cref{fram:Vir}, we apply causal inference techniques to the generated data to identify the relationships between these variables. This process begins by visualizing the data of the $n^{\text{th}}$ versus the $(n+1)^{\text{th}}$ generation virulence (and similarly for transmission) to assess the possibility of causal relationships (see \cref{fig:tSerBoth}).

In \cref{fig:muTser}, the virulence data ($\mu$) exhibits a strong relationship between $\mu_n$ and $\mu_{n+1}$, indicating that the dynamics of virulence can be effectively captured with a lower-dimensional model. This suggests that the relationship between $\mu_n$ and $\mu_{n+1}$ is relatively straightforward, with the data points aligning closely along a one-dimensional manifold. In contrast, \cref{fig:btSer} depicts the transmission rate data ($\beta$), which does not exhibit a clear relationship between $\beta_n$ and $\beta_{n+1}$. The transmission data appears to occupy a higher-dimensional space, indicating more complex dynamics that are not easily captured by its own past values alone. This lack of a clear pattern suggests that additional factors or more sophisticated modeling approaches may be necessary to fully understand the transmission dynamics, highlighting the need to incorporate causal variables into the model. 

Consequently, we test virulence as a potential causal variable for the transmission rate. To visualize this relationship, we plot $\beta_{n+1}$ against $\mu_n$ and $\beta_n$ (see \cref{fig:munBetanBetaP}). The data appear to lie on a two-dimensional manifold, suggesting that the transmission rate at the $(n+1)^{\text{th}}$ generation can be fully explained by the virulence and transmission rate at the $n^{\text{th}}$ generation. This observation underscores that, according to our data, virulence is a Granger causal variable for the transmission rate. A more in-depth analysis can be conducted by following the framework outlined in \cref{fram:Vir} to measure causal inference, thereby demonstrating the key concepts of this project.

\subsubsection{Testing causal variables for virulence in synthetic data}\label{sec:causVi}
Using the available data, we first follow the second step of the framework outlined in \cref{fram:Vir} to understand the causal model for the dataset. In this section, we examine the effect of virulence at a given time, based on its value from the previous time step. This approach can be used to assess whether additional variables, such as the transmission rate, are necessary to accurately predict virulence behavior. To evaluate this potential Granger causality, various causal inference methods can be employed, such as geometric causal inference, transfer entropy, or Granger's ARMA test. In this study, we adopt geometric causal inference, which provides a more detailed analysis from a geometric perspective. Consequently, the dimensionality discussion, supported by the visual tools from the previous section (\cref{sec:DVvbeta}), can now be quantified using the correlation dimension, thereby aiding in the identification of causal variables.

To illustrate the flow of information, specifically the causal effect from the transmission rate ($\beta$) to the virulence ($\mu$), we compute the correlation dimension for multiple time series datasets. The resulting values are presented in \cref{tab:Viru}, and the geometric causal inference ($GeoC_{\beta \to \mu}$) is defined as follows: 
\begin{equation} 
GeoC_{\beta -> \mu}=GeoC(\mu_{n+1}|\mu_n)-GeoC(\mu_{n+1}|\mu_n, \beta_n)=0.000021\approx 0.
\end{equation} 
This result indicates that there is no discernible causal effect of the transmission rate on virulence in the given dataset.
\begin{table}[!ht]
\centering
\renewcommand{\arraystretch}{1.25}
\caption{Correlation dimension $D_2(.)$ and conditional geometric information flow $GeoC(.|.)$ for the virulence $\mu$ and transmission rate $\beta$ datasets, used to compute the geometric causal inference from transmission rate to virulence. Correlation dimension estimates were computed using MATLAB's \texttt{correlationDimension} function \cite{matlabCorrelationDimension}. Estimation procedures for geometric information flow are described in the Methods section.}
\label{tab:Viru}
\begin{tabular}{@{} p{3cm}p{2cm}p{8cm} @{}}
  \toprule
  \textbf{Term} & \textbf{Value}  & \textbf{Description/Remarks}\\
  \midrule $D_2(\mu_n)$  & $0.950295$ & %Reside on a one dimensional manifold.
  \\ 
   $D_2([\mu_n,\mu_{n+1}])$  & $0.950337$ & %Reside on a one dimensional manifold.
   \\ \midrule[0.1pt]
   $GeoC(\mu_{n+1}|\mu_n)$ & $0.000042$  & There are no effective causal variables for $\mu$, and $\mu_{n+1}$ can be modeled by the function $f(\mu_n)$. \\
\bottomrule
\noalign{\vskip 1mm}
\hline
$D_2([\mu_n,~ \beta_n])$  & $1.649863$ & %Reside on a two dimensional manifold.
\\ %\hdashline%\midrule[0.01pt]
   $D_2([\mu_n, ~\mu_{n+1},~ \beta_n])$  & $1.649884$ & %Reside on a two dimensional manifold.
   \\ \midrule[0.1pt]
   $GeoC(\mu_{n+1}|\mu_n,~ \beta_n)$ & $0.000021$  & The inclusion of $\beta_n$ does not provide any additional information. \\
\bottomrule
\noalign{\vskip 1mm}
\hline
\end{tabular}
\end{table}

\subsubsection{Testing causal variables for transmission in synthetic data}\label{Sec:BetaCas}
In this subsection, we focus on identifying the causal variables influencing the transmission rate, utilizing the same synthetic dataset employed in \cref{sec:causVi}. The computed values of the correlation dimension, along with the geometric information measures, are summarized in \cref{tab:trans}. 
\begin{table}[!ht]
\centering
\renewcommand{\arraystretch}{1.25}
\caption{Correlation dimension $D_2(.)$ and conditional geometric information flow $GeoC(.|.)$ for the virulence $\mu$ and transmission rate $\beta$ datasets, used to compute the geometric causal inference from virulence to transmission rate. Correlation dimension estimates were computed using MATLAB's \texttt{correlationDimension} function \cite{matlabCorrelationDimension}.}
\label{tab:trans}
\begin{tabular}{@{} p{3cm}p{2cm}p{8cm} @{}}
  \toprule
  \textbf{Term} & \textbf{Value}  & \textbf{Description/Remarks}\\
  \midrule $D_2(\beta_n)$  & $0.986140$ & Reside on a one-dimensional manifold.\\ %\hdashline%\midrule[0.01pt]
   $D_2([\beta_n,~\beta_{n+1}])$  & $1.660012$ & %Reside on a two-dimensional manifold.
   \\ \midrule[0.1pt]
   $GeoC(\beta_{n+1}|\beta_n)$ & $0.6739$  & There is an effective causal variable for $\beta$, and $\beta_{n+1}$ cannot be modeled solely by $\beta_n$. \\
\bottomrule
\noalign{\vskip 1mm}
\hline
$D_2([\beta_n,~ \mu_n])$  & $1.649863$ & Reside on a two-dimensional manifold.\\ %\hdashline%\midrule[0.01pt]
   $D_2([\beta_n,~ \beta_{n+1},~ \mu_n])$  & $1.649891$ & Reside on a two-dimensional manifold.\\ \midrule[0.1pt]
   $GeoC(\beta_{n+1}|\beta_n,~ \mu_n)$ & $0.000028$  & $\beta_{n+1}$ can be modeled as a function of $\beta_n$ and $\mu_n$. \\
\bottomrule
\noalign{\vskip 1mm}
\hline
\end{tabular}
\end{table}

Notably, while $\beta_n$ lies on a one-dimensional manifold, $[\beta_n,~\beta_{n+1}]$ resides on a two-dimensional manifold. Therefore, additional information is required to fully comprehend the transmission rate. In this case, we tested $\mu_n$ as the additional causal variable influencing the transmission rate. Our findings indicate that both $[\beta_n,\ \mu_n]$ and $[\beta_{n+1},\ \beta_n,\ \mu_n]$ lie on two-dimensional manifolds with approximately the same correlation dimension. This leads us to conclude that virulence is a causal variable for the transmission rate in this dataset. Furthermore, we can compute the geometric causal inference of virulence on the transmission rate, as demonstrated in the following equation.
\begin{align}
        GeoC_{\mu -> \beta}=GeoC(\beta_{n+1}|\beta_n)-GeoC(\beta_{n+1}|\mu_n, \beta_n)=0.673872.
\end{align}

To summarize the overall causal inference discussed in \cref{sec:CausalModel}, we illustrate the causal model of virulence and transmission rate based on the synthetic data in supporting information Figure S9, highlighting the key relationships among the variables.

\subsection{Case study II: LETR analysis of causality and invariant distributions in a modern pandemic dataset}\label{sec:CovidMain}
 
 In the previous example, we highlight the relevance of our approach by using simulated and synthetic data sets that model real-world phenomena. Next, we analyze the transmission and virulence dynamics of a modern pandemic, SARS-CoV-2, using real-time data.

\paragraph{Global pandemic dataset.} For modern pandemic analyses, we use the daily time series of global SARS-CoV-2 case incidence and case fatality rates, obtained from Our World in Data. The dataset covers the period from 1 September 2020 to 24 August 2025. We use the case fatality rate (CFR) \cite{Mathieu2020mortality} as a proxy for virulence \(\mu\) and daily new cases per million people \cite{Mathieu2020cases} as a proxy for transmission. To prevent feature dominance and improve model performance \cite{ozsahin2022impact}, we rescale the transmission series to the unit interval \([0,1]\) using min-max normalization (each data point \(x\) is rescaled as \( \frac{x - \min x}{\max x - \min x}\)) and denote the scaled values by \(\beta\). These measures are coarse proxies that do not necessarily capture within-host virulence or the instantaneous transmission rate. Any alternative proxy that better represents the underlying trait may be substituted into the LETR framework. In this study, we use these proxies to illustrate the practical application of LETR. The primary objective is to assess the plausibility of Granger-style causal drivers for each trait and to test whether one trait exerts a causal influence on the other.
\begin{figure}[ht!]
    \centering
    \begin{subfigure}[b]{0.48\textwidth}
         \centering
\includegraphics[width=\textwidth]{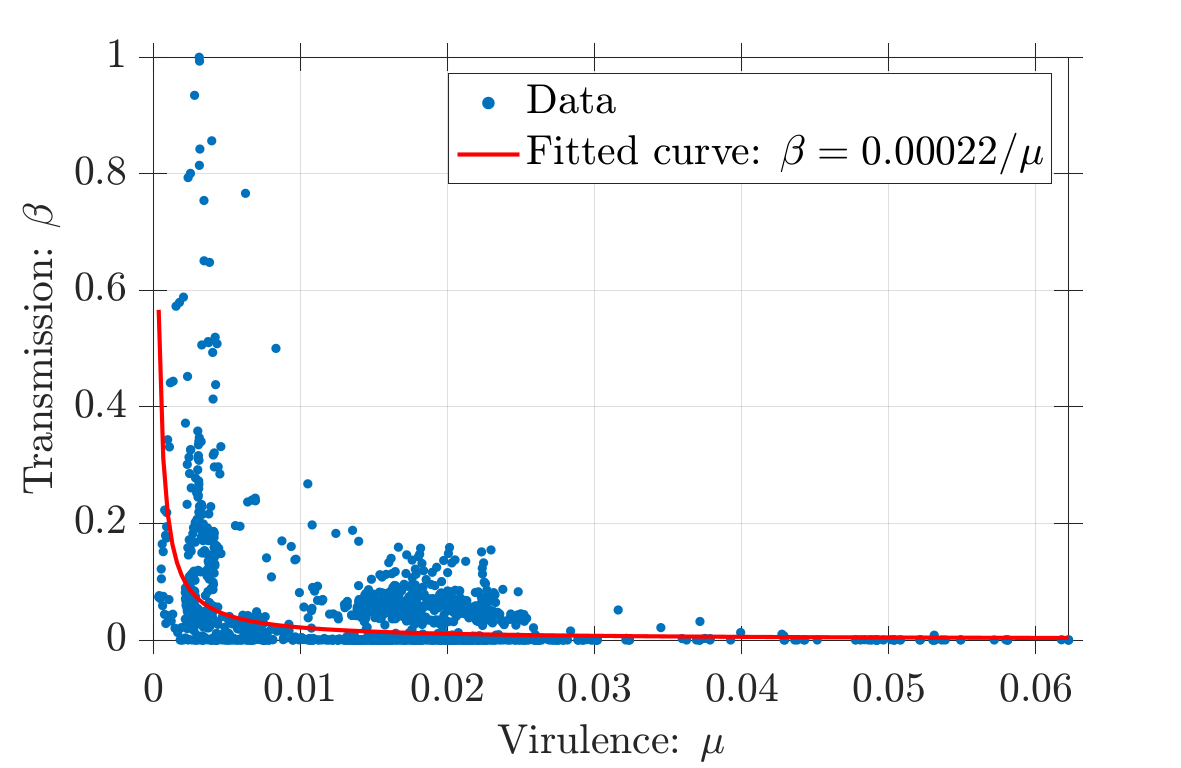}
    \caption{}
    \label{fig:CovidMuBeta}
     \end{subfigure}

         \begin{subfigure}[b]{0.48\textwidth}
             \centering
       \includegraphics[width=\linewidth]{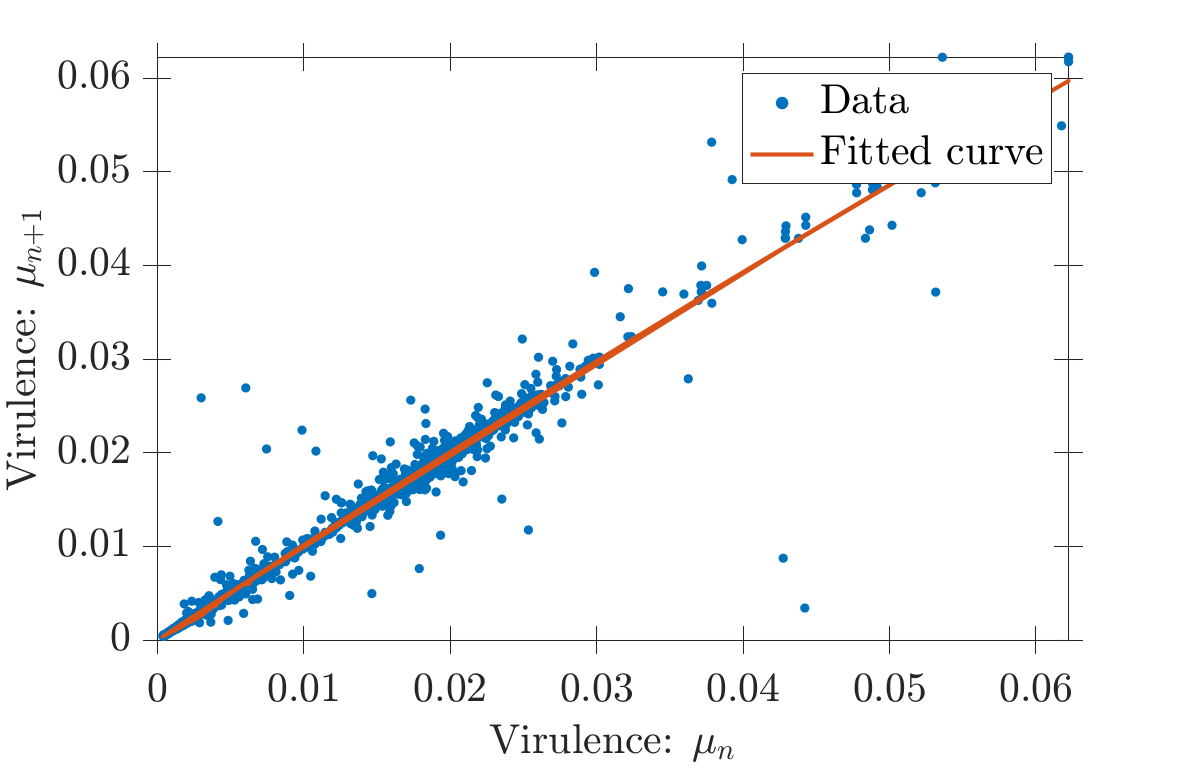}
    \caption{}
    \label{fig:CovidMuNMuNp1}
         \end{subfigure}
         \hfill
                  \begin{subfigure}[b]{0.48\textwidth}
             \centering
       \includegraphics[width=\linewidth]{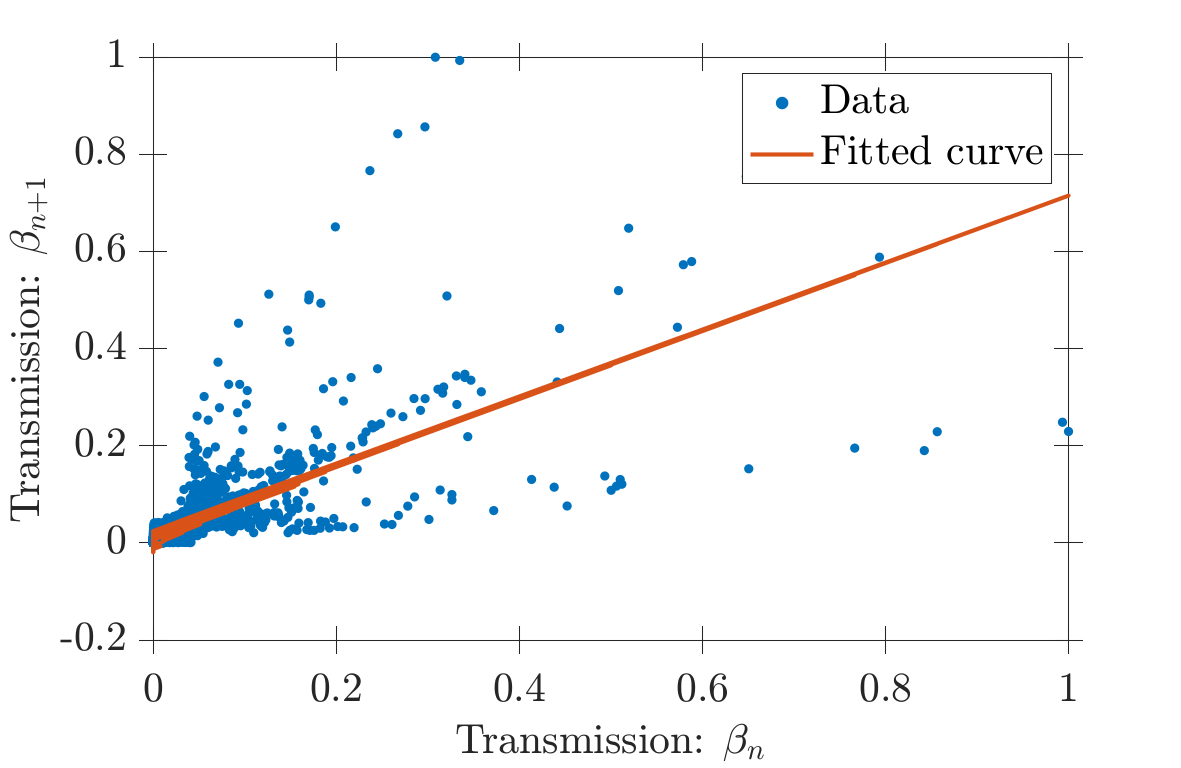}
    \caption{}
    \label{fig:CovidbetaNbetaNp1}
         \end{subfigure}
     
\caption{ \textbf{Geometric investigation of SARS-CoV-2 trait dynamics.}
One-step relationships show that virulence is largely self-predictive while transmission is more dispersed and likely requires additional causal predictors. Data points are daily observations from Our World in Data \cite{Mathieu2020mortality,Mathieu2020cases} for the period 1 September, 2020 to 24 August, 2025. (a) The phase space \(\mu\) versus \(\beta\) with a reciprocal fit \(\beta \approx 0.228/\mu\) obtained by nonlinear regression. (b) shows \(\mu_n\) versus \(\mu_{n+1}\) and reveals a tight one-dimensional manifold consistent with strong temporal predictability of virulence. (c) shows \(\beta_n\) versus \(\beta_{n+1}\), which exhibits substantially greater dispersion and a structured pattern in which higher transmission values separate into three apparent branches, suggesting more complex dynamics and the plausibility of additional causal drivers. Together, these patterns geometrically illustrate the asymmetric causal structure between virulence and transmission and indicate that predicting transmission may require additional ecological, social, or behavioral variables beyond virulence alone.}
    \label{fig:COVIDmunBetan}
\end{figure}

To illustrate the underlying dynamics and the relationship between virulence and transmission, we plot the phase space of virulence versus transmission in \cref{fig:CovidMuBeta} and the corresponding time series in supporting information Figure S1. The scatter plot in \cref{fig:CovidMuBeta} reveals a nonlinear association broadly consistent with a virulence transmission tradeoff and well approximated by an L-shaped inverse curve. Using the MATLAB nonlinear regression function \texttt{nlinfit}, we fit a simple reciprocal model to the data and obtained \(\beta \approx 0.228/\mu\), as shown in \cref{fig:CovidMuBeta}. In this pattern, transmission is high at very low virulence, declines rapidly as virulence increases slightly, and remains low at higher virulence values. At the lower end of the virulence range, the point cloud is highly scattered around the fitted curve. This dispersion may reflect heterogeneity in behavioral, social, and public health responses when measured virulence is low, together with reporting and detection biases that are more influential in that regime. 

\subsubsection{Causal discovery of virulence\textendash transmission dynamics in SARS-CoV-2}\label{sec:causal_covid}
Geometric information flow can be used to identify plausible Granger-style causal drivers of virulence and transmission from their respective time series. In particular, visualization of pairwise time series in scatter plots aids the assessment of these causal hypotheses by revealing low-dimensional structure and predictive geometry that indicate whether one variable carries information about the future of another. \cref{fig:COVIDmunBetan} illustrates these scatter plots. \cref{fig:CovidMuNMuNp1} shows the one-step relationship \(\mu_n\) versus \(\mu_{n+1}\) for virulence and \cref{fig:CovidbetaNbetaNp1} shows the one-step relationship \(\beta_n\) versus \(\beta_{n+1}\) for transmission. These data indicate that virulence, shown in \cref{fig:CovidMuNMuNp1}, is tightly clustered around a clear one-dimensional curve. Therefore, most predictive information on the next time step appears to be contained in the current value of virulence alone. The minor scatter around the curve may reflect observational noise, such as reporting errors or the influence of weak additional drivers. By contrast, the transmission scatter plot in \cref{fig:CovidbetaNbetaNp1} exhibits substantially greater dispersion, although it also shows structure with higher transmission values that separate into three apparent branches. This pattern implies more complex transmission dynamics and suggests the plausibility of additional causal drivers. Therefore, we apply the geometric information flow measure to quantify these hypotheses and to test whether virulence and transmission exert causal effects on one another.

\begin{table}[!ht]
\centering
\renewcommand{\arraystretch}{1.25}
\caption{Geometric information measures for virulence \(\mu\) and transmission \(\beta\) time series for SARS-CoV-2. \(D_2(\cdot)\) denotes the correlation dimension, and \(\mathrm{GeoC}(\cdot\mid\cdot)\) denotes the conditional geometric information flow. Correlation dimension estimates were computed using MATLAB's \texttt{correlationDimension} function \cite{matlabCorrelationDimension}. Geometric information measures indicate that transmission exerts a weak causal influence on virulence, whereas virulence exerts a relatively stronger causal influence on transmission.}
\label{tab:CovidViruTran}
\begin{tabular}{@{} p{3cm}p{2cm}p{11cm} @{}}
  \toprule
  \textbf{Term} & \textbf{Value}  & \textbf{Description/Remarks}\\
  \midrule $D_2(\mu_n)$  & $0.960042$ & %Reside on a one dimensional manifold.
  \\ 
   $D_2([\mu_n,\mu_{n+1}])$  & $1.072071$ & %Reside on a two dimensional manifold.
   \\ \midrule[0.1pt]
   $GeoC(\mu_{n+1}|\mu_n)$ & $0.112029$  &  Low value and scatter plot (\cref{fig:CovidMuNMuNp1}) imply most predictive information for \(\mu_{n+1}\) resides in \(\mu_n\), so \(\mu_{n+1}\) can be modeled as \(f(\mu_n)\). Minor deviations may reflect weak causal influences or observational noise.
  \\
\hline
%\noalign{\vskip 0.01mm}
%\hline
$D_2([\mu_n,~ \beta_n])$  & $1.271778$ & %Reside on a two dimensional manifold.
\\ %\hdashline%\midrule[0.01pt]
   $D_2([\mu_n, ~\mu_{n+1},~ \beta_n])$  & $1.371026$ & %Reside on a two dimensional manifold.
   \\ 
   \midrule[0.1pt]
   $GeoC(\mu_{n+1}|\mu_n,~ \beta_n)$ & $0.099248$  & Including \(\beta_n\) provides only marginal additional predictive information beyond \(\mu_n\). \\
\midrule[1pt]
\multicolumn{3}{c}{$GeoC_{\beta -> \mu}=GeoC(\mu_{n+1}|\mu_n)-GeoC(\mu_{n+1}|\mu_n, \beta_n)=0.012781$}\\
%\noalign{\vskip 1mm}
\midrule[1pt]
$D_2(\beta_n)$  & $0.693734$ & %Reside on a one dimensional manifold.
\\ %\hdashline%\midrule[0.01pt]
   $D_2([\beta_n,~\beta_{n+1}])$  & $1.001665$ & %Reside on a two-dimensional manifold.
   \\ \midrule[0.1pt]
   $GeoC(\beta_{n+1}|\beta_n)$ & $0.307931$  & There may be an effective causal variable for $\beta$, and $\beta_{n+1}$ may not be modeled solely by $\beta_n$. \\
\bottomrule
%\noalign{\vskip 1mm}
%\hline
$D_2([\beta_n,~ \mu_n])$  & $1.271778$ & %Reside on a two dimensional manifold.
\\ %\hdashline%\midrule[0.01pt]
   $D_2([\beta_n,~ \beta_{n+1},~ \mu_n])$  & $1.309606$ & %Reside on a two dimensional manifold.
   \\ \midrule[0.1pt]
   $GeoC(\beta_{n+1}|\beta_n,~ \mu_n)$ & $0.037828$  & Virulence shows a detectable causal influence on transmission, but additional variables may be required to fully account for transmission dynamics.
  \\
\midrule[1pt]
\multicolumn{3}{c}{$GeoC_{\mu -> \beta}=GeoC(\beta_{n+1}|\beta_n)-GeoC(\beta_{n+1}|\mu_n, \beta_n)=0.270103$}\\
%\noalign{\vskip 1mm}
\midrule[1pt]
\noalign{\vskip 0.1mm}
\hline
\end{tabular}
\end{table}

\cref{tab:CovidViruTran} reports correlation dimension estimates and conditional geometric information flow measures for the SARS-CoV-2 virulence and transmission time series. The correlation dimension of virulence alone, \(D_{2}(\mu_{n}) \approx 0.96\), is close to \(D_{2}([\mu_{n},\mu_{n+1}]) \approx 1.07\). Together with \(\mathrm{GeoC}(\mu_{n+1}\mid\mu_{n}) \approx 0.112\), these values indicate that most short-term predictability of \(\mu_{n+1}\) is contained in \(\mu_{n}\). Including \(\beta_{n}\) provides only marginal additional information, with \(\mathrm{GeoC}(\mu_{n+1}\mid\mu_{n},\beta_{n}) \approx 0.099\) and a small net effect \(GeoC_{\beta\to\mu}\approx 0.013\). This pattern suggests that transmission is, at best, a weak causal predictor of subsequent virulence. By contrast, transmission exhibits a different profile. The one-step predictability of \(\beta\) from its own past, \(\mathrm{GeoC}(\beta_{n+1}\mid\beta_{n}) \approx 0.308\), suggests that additional co-variates are needed to model transmission dynamics. When \(\mu_{n}\) is incorporated, the conditional measure falls to \(\mathrm{GeoC}(\beta_{n+1}\mid\beta_{n},\mu_{n}) \approx 0.038\), yielding a net contribution \(GeoC_{\mu\to\beta}\approx 0.270\). This result indicates that virulence supplies substantial complementary information for predicting future transmission. Taken together, these quantitative findings formalize and support the asymmetry visible in the scatter plots. In other words, virulence has a detectable causal influence on subsequent transmission, while transmission provides only weak predictive information for future virulence. These inferences remain conditional on the quality and resolution of the observational data and are subject to potential biases from reporting errors, unobserved confounders and other sources of noise, which motivate follow up analyses with higher resolution co-variates and experimental validation.

\paragraph{Regional comparison of causal measures}
Applying LETR to the global SARS-CoV-2 time-series data revealed a particular set of relationships between virulence and transmission. But global pandemics are often defined by regional variation in their dynamics, which can be traced to any number of factors that differ from setting to setting. To assess whether the asymmetry observed in the global series is consistent across spatial (geographical) scales, we computed geometric information flows for three regional aggregates, and report the summary measures in \cref{tab:summaryRegions}. Specifically, the causal influence of transmission on virulence is uniformly weak across all scales ($GeoC_{\beta\to\mu}= 0.0128$ globally and $0.0711$ - $0.0919$ regionally), whereas the causal influence of virulence on transmission varies substantially, being comparatively strong at the global ($GeoC_{\mu\to\beta}=0.2701$) and Asian ($0.2460$) scales but very weak in the USA ($0.0246$) and Europe ($0.0420$). The table indicates that virulence provides substantially more predictive information on future transmission at the global and Asian scales. The USA and Europe show comparatively smaller effects of virulence on transmission. In contrast, the causal influence of transmission on virulence remains weak across regions and is consistent with the global pattern. These regional differences may reflect variation in virus evolution, epidemic timing, behavior, socioeconomic status, reporting practices, or unobserved covariates \cite{pan2024assessing,bourdin2023regional}. Detailed computational results for each region are reported in Section~S1 of the Supplementary Information. Additional visualizations and supporting analyses are provided in the Supplementary Information.

\begin{table}[ht]
\centering
\caption{Summary of geometric information measures at regional and global scales. Values shown are $GeoC_{\beta\to\mu}$ (causal influence of transmission on virulence) and $GeoC_{\mu\to\beta}$ (causal influence of virulence on transmission). Lower values indicate weaker causal influence. Overall, the causal influence of transmission on virulence is markedly weaker. In the USA and Europe, virulence exerts only a very weak causal influence on transmission, whereas at the global and Asian scales virulence exerts a comparatively stronger influence. Full numerical results are provided in the Supplementary Information.}
\label{tab:summaryRegions}
\begin{tabular}{lcc}
\toprule
Region & $GeoC_{\beta\to\mu}$ & $GeoC_{\mu\to\beta}$ \\
\midrule
Global & 0.0128 & 0.2701 \\
USA    & 0.0918 & 0.0246 \\
Europe & 0.0919 & 0.0420 \\
Asia   & 0.0711 & 0.2460 \\
\bottomrule
\end{tabular}
\end{table}

\subsubsection{Estimating invariant trait distributions for SARS-CoV-2 virulence and transmission}
Generative maps for virulence and transmission may be inferred by regression. In this study, we fit one-step generative functions using the least absolute shrinkage and selection operator (LASSO) \cite{tibshirani1996regression} and overlay the fitted maps on the data in \cref{fig:CovidMuNMuNp1,fig:CovidbetaNbetaNp1}. Because the chosen observables may not span the full causal state space, the fitted maps do not always capture all observed variability. Nonetheless, these maps are useful for short-term prediction and for formulating mechanistic hypotheses about trait updates. When the fitted map displays sensitive dependence on initial conditions, or when stochastic forcing is important, individual trajectories become unpredictable while the population-level distribution of trait values remains informative. In such cases, we estimate the transfer operator directly from data and derive long-term behavior from its invariant density. Practical, data-driven estimators include Ulam's method \cite{ulam1960problems} and kernel-based estimators \cite{surasinghe2024learning}, which both approximate the Frobenius Perron operator and whose leading eigenvector yields the invariant density.

\begin{figure}[ht!]
    \centering
    \begin{subfigure}{0.48\textwidth}
        \centering
        \includegraphics[width=\textwidth]{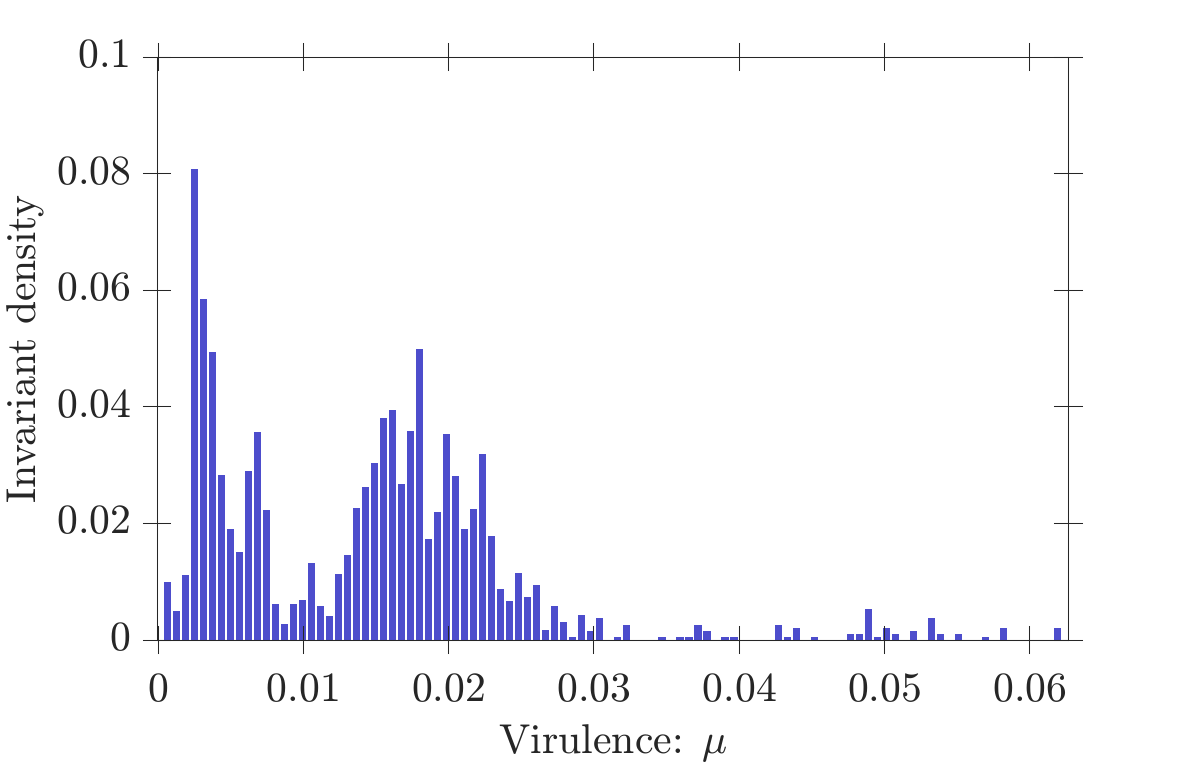}
        \caption{}
        \label{fig:CovidMuInvDen}
    \end{subfigure}
    \begin{subfigure}{0.48\textwidth}
        \centering
    \includegraphics[width=\textwidth]{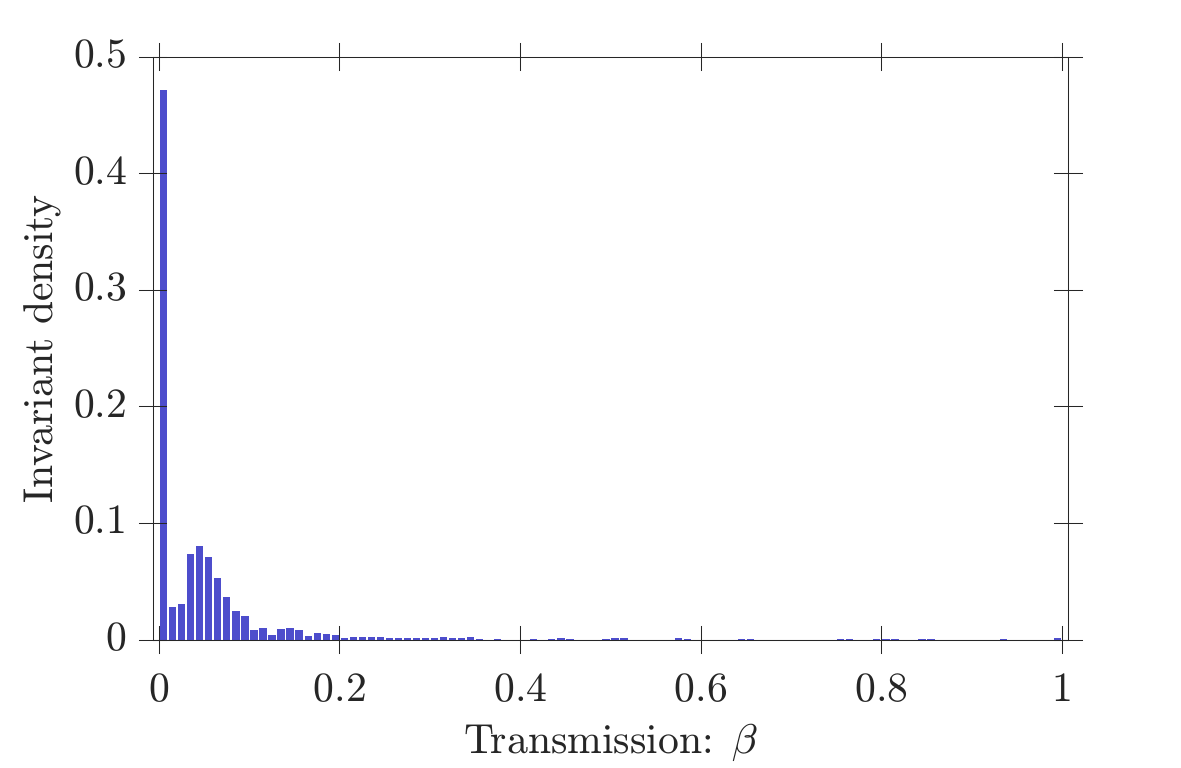}
        \caption{}
        \label{fig:CovidBetaInvDen}
    \end{subfigure}
    
\caption{\textbf{Long-run trait distributions of SARS-CoV-2.}
Estimated invariant densities for SARS-CoV-2 virulence and transmission obtained by Ulam's method. (a) The invariant density for virulence \(\mu\) is bimodal with modes near \(\mu\approx 0.0025\) and \(\mu\approx 0.0175\). (b) The invariant density for transmission \(\beta\), where \(\beta\) denotes the min-max scaled (see \cite{jain2011min,ali2022investigating}) daily new cases per million. The transmission density is strongly concentrated near the lower bound and displays a rapidly decaying tail that is approximately exponential, with a weak secondary mode at the very low end. Data are daily observations from Our World in Data \cite{Mathieu2020mortality,Mathieu2020cases} covering the period from September 1st, 2020, to August 24, 2025. These estimates indicate a long-term tendency toward low virulence and low transmission. The observed bimodality in virulence may arise from variations in periods favorable to the pathogen, driven by host heterogeneities or environmental fluctuations.}

    \label{fig:COVIDinvDen}
\end{figure}

Estimated invariant densities computed by Ulam's method are shown in \cref{fig:CovidBetaInvDen}. The invariant density for virulence, plotted in \cref{fig:CovidMuInvDen}, is bimodal with modes near \(\mu\approx 0.0025\) and \(\mu\approx 0.0175\). This pattern indicates that virulence values tend to concentrate at very low and slightly elevated levels in the long run. Possible explanations for this bimodality include the coexistence of distinct viral strains, heterogeneous vaccine coverage, or other ecological factors. The invariant density for transmission, shown in \cref{fig:CovidBetaInvDen}, is strongly concentrated near the lower end of the support and exhibits a rapidly decaying tail that resembles an exponential distribution for larger values. A weak secondary mode is apparent at the very low end of the transmission range. Taken together, the estimated densities suggest that, for the period and observables analyzed here, both virulence and transmission are more likely to occupy low values in the long term. A summary of the key insights from both the synthetic myxomatosis dataset and the SARS-CoV-2 case study, together with their broader implications for infectious disease research, is provided in \cref{tab:combined_aspects}.

\begin{table}[ht!]
%\centering
\renewcommand{\arraystretch}{1.25}
\small
\caption{Comparison of key elements of LETR across the synthetic benchmark (myxomatosis-motivated) and SARS-CoV-2 applications, highlighting shared and distinct causal patterns in virulence–transmission dynamics.}
\label{tab:combined_aspects}
\begin{tabular}{@{} p{3.2cm} p{4.8cm} p{4.8cm} p{5.0cm} @{}}
\toprule
\textbf{Element} & \textbf{Case study I: Synthetic benchmark (myxomatosis-motivated)} & \textbf{Case study II: pandemic application} & \textbf{Implications for disease evolution and ecology} \\
\midrule
\textbf{Predictive relationship} &
LETR recovers the imposed directional influence and demonstrates that virulence drives transmission. &
In SARS-CoV-2 data, the inferred direction varies by scale and setting. In some aggregates virulence precedes transmission, while elsewhere the effect is very weak or absent. This pattern likely reflects biological mechanisms and setting specific factors such as epidemic phase, behavior, testing and reporting. Causal claims therefore require local diagnostics. & Virulence $\rightarrow$ transmission is dynamically asymmetric and scale-dependent. Monitoring virulence can improve short term forecasts in some settings but inference requires corroboration with local data on behavior, testing and reporting. \\
\midrule
\textbf{Trait interdependence} &
Geometric diagnostics capture how traits co-vary in synthetic systems. &
In this regime, transmission and virulence alone do not fully explain the behavior of epidemics. &
Indicates that complex trait coupling demands multi-trait models. \\
\midrule
\textbf{Long-term distributions} &
Simulations produce chaotic attractors for virulence and transmission consistent with benchmark design.  &
Virulence shows a bimodal distribution (modes near 0.0025 and 0.0175); transmission is low with an exponential-like tail. &
Reveals distinct evolutionary or ecological regimes influencing virulence outcomes. \\
\midrule
\textbf{Possible mechanisms} &
Framework recovers known mechanistic directionality from imposed system. &
Peaks in virulence may reflect different viral strains or host-specific responses. &
Highlights capacity to identify mechanistic heterogeneity from observational data. \\
\midrule
\textbf{Analytical contribution} &
LETR pipeline demonstrates causal discovery using geometric diagnostics. &
“Invariant-density” estimates provide compact summaries of long-term behavior. &
Integrates causal inference with dynamical-systems diagnostics for predictive epidemiology. \\
\bottomrule
\end{tabular}
\end{table}

\section{Discussion}

This study proposes \emph{Granger‐causality philosophy} as the basis of our proposed \emph{learning evolutionary trait relationships} (LETR) framework. Contemporary causal inference is dominated by two complementary schools. One is the predictive, time-series oriented approach, originating from the Nobel Prize-winning work of Clive W.J. Granger \cite{NobelGren2004}, which tests whether past observations of one variable improve forecasts of another \cite{granger1969investigating,Wiener1956}. The other is the interventionist or counterfactual approach epitomized by Pearl which formalizes causation in terms of hypothetical manipulations and do-calculus \cite{pearl2010causal}. Bayesian network and causal graph methods have become increasingly popular across disciplines and have seen substantial uptake in biological applications \cite{laubach2021biologist,kurth2021continuing,barrero2025causal,shipley1999testing}. However, quantitative traits are naturally observed as time series and Granger style analysis aligns directly with forecasting goals and with the practical question of whether one trait improves the prediction of another \cite{guo2010granger,runge2019inferring,deyle2016global}.

LETR adopts the predictive information flow perspective. It also recognizes the continued value of interventionist and graphical approaches when experiments or explicit identification assumptions are available. In practice, we implement a geometric version of Granger style inference \cite{surasinghe2020geometry} because it provides a visual and quantitative complement to probabilistic measures such as transfer entropy \cite{schreiber2000measuring}. Geometric diagnostics test whether the addition of a candidate predictor reduces the effective dimensionality of the one-step mapping, and thereby supplies information about the future of the focal variable. This approach avoids some technical limitations of entropy-based estimators and is particularly useful when interactions are strongly nonlinear or when low-dimensional structure is present in the data. Geometric methods also facilitate intuitive exploratory analysis and hypothesis generation for downstream, more assumption-heavy Granger causality tests. 

We apply this to a long-standing pillar of disease evolution and ecology: the relationship between virulence and transmission in pathogen-host interactions. Our findings demonstrate that causal structure among pathogen traits is dynamically asymmetric and scale-dependent, generating stable evolutionary attractors that cannot be captured by tradeoff theory or correlational analyses alone.

\paragraph{Case study I: Myxomatosis outbreak-inspired synthetic data} 
In the myxomatosis-inspired system, case mortality modifies within-host titers, which  determine the time course of infectiousness, and the probability of infection from disease vectors (hematophagic insects). This mechanistic chain informed our synthetic design and provided a clear ground truth for testing LETR. Applying LETR to the synthetic data, we identified Granger style predictors with geometric diagnostics and correlation dimension estimates. Geometric causal measures recovered the imposed directional influence from virulence to transmission and confirmed that transmission does not causally drive virulence. The synthetic validation demonstrates two concrete uses of LETR for the empirical myxomatosis problem. Firstly, LETR tests whether observed trait time series are consistent with virulence acting upstream of transmission by asking whether inclusion of virulence substantially reduces the dimension of the predictive manifold and improves one-step forecasts. Second, LETR provides quantitative evidence for directional influence that can be aligned with mechanistic intuition from within host to population level dynamics. Extensions that detect indirect pathways such as causation entropy can be layered into LETR to explicitly test whether within host titer mediates the effect of virulence on transmission \cite{sun2014causation}. Together these steps show how the mechanistic intuition from myxomatosis can be translated into concrete, data driven tests that yield testable epidemiological patterns and insights. The implications of these patterns for disease evolution and ecology are summarized in \cref{tab:combined_aspects}.

\paragraph{Case study II: Modern pandemic using SARS-CoV-2 epidemiological data}  Applying LETR to the global SARS-CoV-2 series, we find that changes in virulence \(\mu\) generally provide more predictive information for subsequent transmission \(\beta\) than the reverse. This asymmetry is strongest at the global scale and in Asia and weaker in the USA and Europe. The information flow from transmission to virulence remains weak across regions. The patterns and their implications for disease evolution and ecology are summarized in \cref{tab:combined_aspects}. Further methodological details and per-region supporting analyses are provided in \cref{sec:causal_covid} and Supplementary Information Section~S1.

Estimated invariant densities indicate long-term tendencies toward low virulence and low transmission. Specifically, virulence is bimodal with modes near \(\mu\approx 0.0025\) and \(\mu\approx 0.0175\), while transmission \(\beta\) is strongly concentrated below 0.2 with an approximately exponential tail \cref{fig:COVIDinvDen}. The bimodality in virulence may reflect the coexistence of distinct viral lineages, or it may arise from host-level heterogeneity, including behavioral, social, and ecological factors that shape exposure and outcomes \cite{barrero2025causal,deyle2016global}. High virulence episodes can also lead to increased public awareness, vaccination, and other interventions that suppress both severity and transmission which aligns with the asymmetric causal pattern we observe \cite{bull1994virulence,fierce2022high,de2008virulence}. Social inequalities and spatial clustering of vulnerable individuals can sustain systematic differences in severity across subpopulations \cite{farmer2016social,muurling2023last,surasinghe2024structural,blackstone2020variation}. Finally, environmental variation that alters the duration of periods favorable to the pathogen can produce a mix of rare severe outbreaks and frequent mild outbreaks, generating a bimodal long-term distribution \cite{anttila2015environmental}.

\paragraph{Implications for the ecology and evolution of infectious disease}

Through simulated and real-world pandemic examples, we observe the peculiar nature of the relationship between virulence and transmission. Far from being a theoretical exercise, the findings offer a careful means of advancing a subversive idea: traits assumed to be the product of evolutionary forces can also be driven by ecological or behavioral factors. This underscores the need to consider the statistical concept of model identifiability---where differing model structures can fit a given epidemiological time-series curve \cite{cortez2013distinguishing, Vajda1988}---when inferring the forces driving virulence, transmission, and the shape of epidemics. That is, one can mistakenly interpret changes in virulence and/or transmission as the result of molecular evolution when those changes very well could have arisen from non-genetic (e.g., ecological) forces. These findings implore us to think carefully about the relationship between virulence and transmission, and reconcile the evolutionary and ecological forces as fundamental (often co-occurring) shapers of disease dynamics. Relatedly, our results highlight the need for fine-grained genetic, clinical, and field data to resolve patterns such as the dual virulence peaks observed for SARS-CoV-2. They also motivate integrating modeling and empirical surveillance with ecological and mechanistic studies to strengthen causal inference between traits of interest.

\paragraph{Conclusions} 

As a general-purpose, data-driven framework, LETR opens new avenues for mechanistic understanding and predictive modeling in the era of large data sets. This allows researchers not only to ask ``which traits drive others,''  but also to answer ``what trait diversity will emerge over time'' (as characterized by the invariant density derived from transfer operators). This unified approach connects mechanistic causality with population-level patterns, bridging the gap between individual-level trait dynamics and macro-scale outcomes.

Although our study focused on the relationship between virulence and transmission, the LETR framework can be extended to other trait pairings. For instance, LETR can be used to identify causal relationships between traits and to quantify how covariates such as ecological context, host demography, and treatment regimes influence these relationships. Potential applications include: antibiotic resistance versus growth rate \cite{ahmad2025role}; host resistance versus tolerance \cite{raberg2007disentangling}; life-history tradeoffs between growth and reproduction \cite{stearns1998evolution}; and phenotypic plasticity versus canalization \cite{fusco2010phenotypic}. Consequently, the same underlying trait pairing can manifest differently under specific conditions \cite{fussmann2007eco}.

\section*{Acknowledgments}
The authors would like to thank the 2024 and 2025 meetings of the Society for Modeling and Theory in Population Biology, where ideas germane to this article were discussed and developed. The authors thank members of the OGPlexus for helpful feedback on various aspects of the project. 

\section*{Funding} This work was supported by the Santa Fe Institute and the Mynoon and Stephen Doro MD, PhD Family Private Foundation Fund.

\section*{Data availability}
Data and code are available on Git: https://github.com/OgPlexus/learningvirulence

\section*{Author Contributions}
SS and CBO conceived the study, developed the models, interpreted the results, and wrote the original and revised versions of the manuscript.

\clearpage
{\small
\printbibliography[title={References}]
}
\end{refsection}

% \clearpage
% {\small 
% \bibliography{refs}}
\clearpage
\begin{refsection}
\counterwithin*{section}{part}
\renewcommand{\thesection}{S\arabic{section}}

\setcounter{figure}{0}
\renewcommand{\thefigure}{S\arabic{figure}}
\setcounter{table}{0}
\renewcommand{\thetable}{S\arabic{table}}

\newcommand{\RegionFigure}[3]{%
  \begin{figure}[htb!]
    \centering
    \begin{subfigure}[b]{0.48\textwidth}
      \centering
      \includegraphics[width=\textwidth]{#1COVIDMuBeta.png}
      \caption{}%
      \label{fig:#2a}
    \end{subfigure}%
    \hfill
    \begin{subfigure}[b]{0.48\textwidth}
      \centering
      \includegraphics[width=\textwidth]{#1COVIDmuNmuNp1.png}
      \caption{}%
      \label{fig:#2b}
    \end{subfigure}

    \vspace{0.6em}

    \begin{subfigure}[b]{0.48\textwidth}
      \centering
      \includegraphics[width=\textwidth]{#1COVIDbetaNbetaNp1.png}
      \caption{}%
      \label{fig:#2c}
    \end{subfigure}

    \caption{#3}
    \label{fig:#2}
  \end{figure}
}
%%%%%%%%%%%%%%%%%%%%%%%%%%%%%%
%%%% SI %%%%%%%%%%%%%%%%%%%%%%
%%%%%%%%%%%%%%%%%%%%%%%%%%%%%%%%
{\huge\noindent\textbf{Supporting Information for} }\\[1ex]
  \maketitle
%\vspace{2em}
 %\section*{This PDF file includes:}
 \renewcommand{\contentsname}{This PDF file includes:}
 \tableofcontents
\clearpage
\section{Results for SARS-CoV-2}
In the main text, we analyze SARS-CoV-2 at the global level. As noted there, the results may be affected by numerous co-variates, which makes the causal relationship between virulence and transmission appear weak in an aggregated setting. This highlights the extent to which these relationships are context-dependent. Many of these co-variates, including environmental conditions, behavioral patterns, public health policies, and socio-cultural factors, can be mitigated when the analysis is conducted at the country or regional level. To illustrate this, the present supporting materials provide parallel analyses for the United States, Europe, and Asia.

\subsection*{Time series of SARS-CoV-2 virulence and transmission}
For completeness, we present the raw time series used in the analysis for both global and selected regional aggregates. The daily case fatality rate (CFR) is used as a proxy for virulence, and transmission is measured by daily new cases per million people rescaled to the unit interval using min–max normalization. The figure displays a global (top) and three compact regional panels (bottom) showing the USA, Europe, and Asia. In the plots, CFR is shown in black and transmission in blue. The series follows the available data range \cite{Mathieu2020mortality,Mathieu2020cases}, and provide empirical input for the subsequent LETR analysis.

\begin{figure}[ht!]
    \centering
    \includegraphics[width=\linewidth]{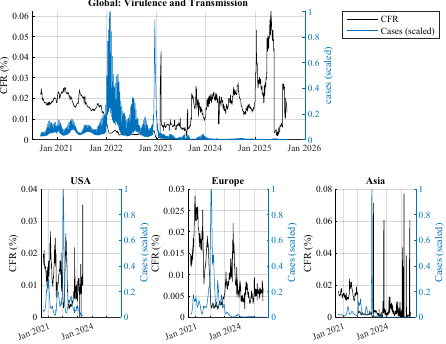}
    \caption{Time series of SARS-CoV-2 virulence and transmission. The top panel shows global CFR (left axis, black) and transmission (right axis, daily new cases per million, min–max scaled to $[0,1]$, blue). The three lower panels present the corresponding regional series for the USA, Europe, and Asia, plotted using the same color convention. Each regional panel uses its own available date range following the data in \cite{Mathieu2020mortality,Mathieu2020cases}.}
    \label{fig:CovidMuBetaTseries}
\end{figure}

\subsection{Region-specific causal analyses}

The following sections present the region-by-region analyses that parallel the global workflow described in the main text. Each regional subsection reproduces the same diagnostic pipeline used for the global data and provides scatterplots, low-dimensional geometry diagnostics, and conditional geometric information flow measures. The objective is to assess whether the asymmetry between virulence and transmission documented at the global level persists at regional scales and to identify region-specific features that may inform mechanistic hypotheses.
\paragraph{Data and proxies}
For each region, we analyze two primary series. Virulence is measured by the daily case fatality rate (CFR), computed as the number of confirmed deaths divided by the number of confirmed cases on the same reporting date unless otherwise stated. Transmission is measured by the daily new confirmed cases per million people. Transmission series are min--max scaled to the unit interval $[0,1]$. Where relevant, each regional panel uses the local date vector for that region \cite{Mathieu2020mortality,Mathieu2020cases}. 

\paragraph{Analytical methods}
The results reported here are obtained by applying the same analytical pipeline described in the main text. In brief:
\begin{itemize}
  \item Pairwise scatterplots of \(\mu_n\) vs \(\mu_{n+1}\) and \(\beta_n\) vs \(\beta_{n+1}\) are used to visually inspect low-dimensional structure and predictive geometry.
  \item Correlation dimension $D_2(\cdot)$ estimates are computed with MATLAB's \texttt{correlationDimension} routine \cite{matlabCorrelationDimension}. 

\item Geometric information flow measures are computed following the procedure described in the Methods. We report both conditional geometric information measures and derived directional geometric information flow quantities. For example, we evaluate $\mathrm{GeoC}(\mu_{n+1}\mid\mu_n)$ and $\mathrm{GeoC}(\mu_{n+1}\mid\mu_n,\beta_n)$, and we compute directional flow measures such as $GeoC_{\beta\to\mu}=\mathrm{GeoC}(\mu_{n+1}\mid\mu_n)-\mathrm{GeoC}(\mu_{n+1}\mid\mu_n,\beta_n)$.

\end{itemize}

\subsubsection{Results of regional causal analysis}
Here, we summarize the regional estimates of correlation dimension and geometric information flow in \cref{tab:GeoC_by_region}, with corresponding phase space and one-step scatterplots shown in \cref{fig:GeoC_USA,fig:GeoC_EU,fig:GeoC_ASIA}. These results allow a direct comparison with the global analysis reported in the main text.

\begin{table}[ht!]
\centering
\small % compact font for high density tables
\renewcommand{\arraystretch}{1.15}
\caption{Selected geometric information measures by region. Values correspond to correlation dimension $D_2$ and conditional geometric information $\mathrm{GeoC}$. Here, $\mathrm{GeoC}_{x\to y}$ denotes the geometric information flow measure used to assess causal influence from $x$ to $y$ as described in Methods section~2.1 of the main text. The $D_2$ estimates were computed with MATLAB's \texttt{correlationDimension} function and parameters were selected by visual inspection of the function's interactive log--log plot of the correlation sum against neighbourhood radius. Reported values are rounded to six decimal places. Overall, the measures indicate that transmission exerts a weak causal influence on virulence at both global and regional scales. For the USA and Europe, virulence also shows a weak causal influence on transmission. By contrast, at the global and Asia scales, virulence exerts a relatively stronger causal influence on transmission.}
\label{tab:GeoC_by_region}
\begin{tabular}{@{}lcccc@{}}
\toprule
\textbf{Term} & \textbf{Global} & \textbf{USA} & \textbf{Europe} & \textbf{Asia} \\
\midrule
$D_2(\mu_n)$ & 0.960042 &  0.950158
              &  0.927558
              &  0.899594
              \\[4pt]
$D_2([\mu_n,\mu_{n+1}])$ & 1.072071 & 1.095456  
                &  1.215081
                & 1.153344 \\[4pt]
$D_2([\mu_n,\beta_{n},\mu_{n+1}])$ & 1.371026 & 1.533408
                &  1.537789
                & 1.337165 \\ [1pt]
                \midrule[0.1pt]
$D_2([\mu_n,\beta_{n}])$ & 1.271778 & 1.479956
                        &  1.342144
                        & 1.154491  \\ [1pt]
                        \midrule[0.1pt]
$D_2(\beta_n)$ & 0.693734 &  0.892532
                &  0.795365
                & 0.737713 \\[4pt]
$D_2([\beta_n,\beta_{n+1}])$ & 1.001665 & 1.074973 
                            &  0.854573
                            & 1.004846 \\[4pt]
$D_2([\mu_n,\beta_{n},\beta_{n+1}])$ &1.309606  &  1.637780
                                    &  1.359345
                                    & 1.188627 \\[4pt]
\midrule[0.1pt]
$GeoC(\mu_{n+1}\mid\mu_n)$ & 0.112029 &  0.145298
                            &  0.287523
                            & 0.25375 \\[4pt]
$GeoC(\mu_{n+1}\mid\mu_n,\beta_n)$ & 0.099248 &  0.053452
                                &  0.195645
                                & 0.182674 \\[4pt]
\midrule[0.1pt]
$GeoC_{\beta\to\mu}$ & 0.012781 & 0.091846
                    &  0.091878
                    & 0.071076 \\
\bottomrule
\noalign{\vskip 1mm}
\hline
$GeoC(\beta_{n+1}\mid\beta_n)$ & 0.307931 & 0.182441  
                                &  0.059208
                                & 0.267133 \\[4pt]
$GeoC(\beta_{n+1}\mid\beta_n,\mu_n)$ & 0.037828 &  0.157824
                                    &  0.017201
                                    & 0.021098 \\[4pt]
\midrule[0.1pt]
$GeoC_{\mu\to\beta}$ & 0.270103 &  0.024617
                    &  0.042007
                    & 0.246035 \\
\bottomrule
\noalign{\vskip 1mm}
\hline
\end{tabular}
\end{table}

%\subsubsection{USA}

% Usage in main text: USA figure, high impact journal style
\RegionFigure{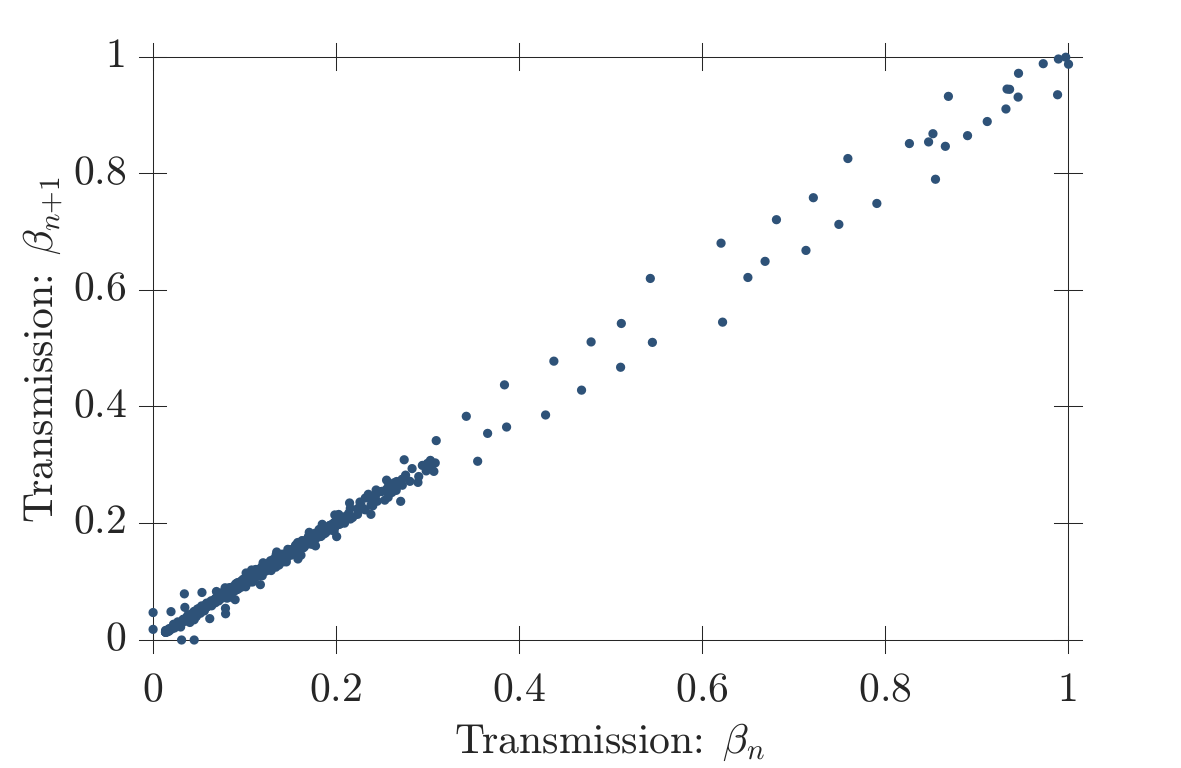}{GeoC_USA}{\textbf{Geometric investigation of SARS-CoV-2 trait dynamics for the United States.} One-step relationships show that both virulence and transmission are largely self-predictive. The USA data are daily observations from Our World in Data \cite{Mathieu2020mortality,Mathieu2020cases} covering the period from September 1st, 2020, to May  22nd, 2023. (a) shows the joint phase space of virulence \(\mu\) and transmission \(\beta\). (b) shows \(\mu_{n+1}\) plotted against \(\mu_n\). The horizontal axis shows \(\mu_n\) and the vertical axis shows \(\mu_{n+1}\), which emphasizes short-term predictability of virulence. (c) shows \(\beta_{n+1}\) plotted against \(\beta_n\). In contrast to the global transmission data (see main text section 3.2.1, Figure 4c), where higher transmission values separate into three apparent branches, the USA transmission series lacks that branching. This observation suggests a reduced role for additional causal drivers of transmission at the national level.}

%\subsubsection{EU}

% Usage in main text: Europe figure, high impact journal style
\RegionFigure{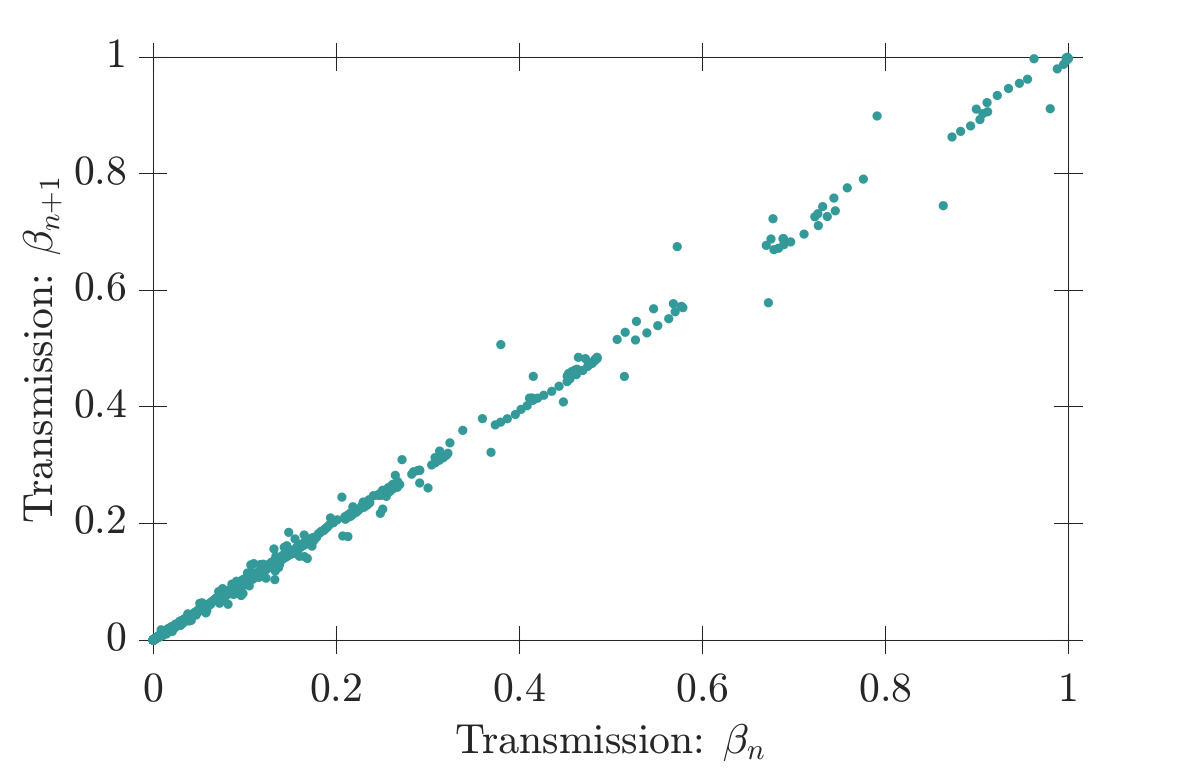}{GeoC_EU}{\textbf{Geometric investigation of SARS-CoV-2 trait dynamics for Europe.} One-step relationships show that both virulence and transmission are largely self-predictive. The European data are daily observations from Our World in Data \cite{Mathieu2020mortality,Mathieu2020cases} covering a period from September 1st, 2020, to August 24, 2025. (a) shows the joint phase space of virulence \(\mu\) and transmission \(\beta\). (b) shows \(\mu_{n+1}\) plotted against \(\mu_n\), and (c) shows \(\beta_{n+1}\) plotted against \(\beta_n\). As with the USA data, the European transmission series does not display the three-branch pattern reported for global transmission in the main text section 3.2.1 Figure 4c. This similarity points to a reduced role for additional causal drivers of transmission at these regional scales.}

%\subsubsection{Asia}
% Usage in main text: Asia figure, concise caption
\RegionFigure{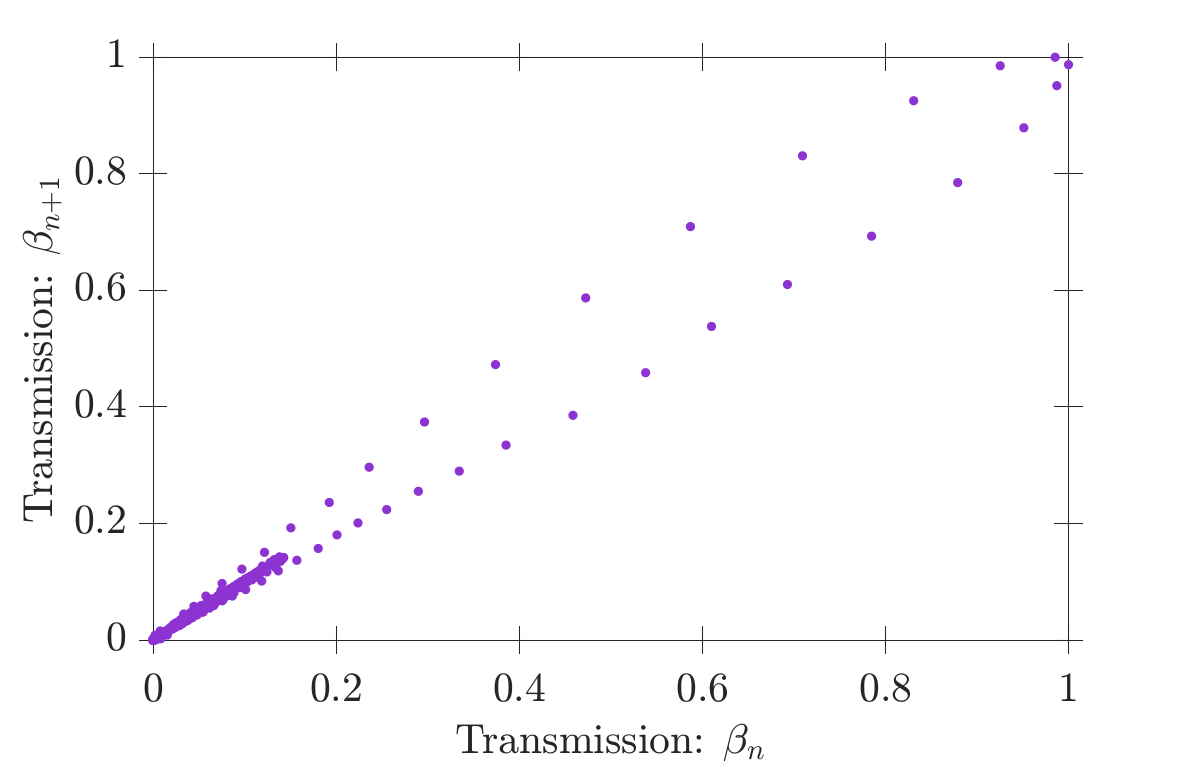}{GeoC_ASIA}{\textbf{Geometric investigation of SARS-CoV-2 trait dynamics for Asia.} One-step relationships indicate that virulence and transmission are largely self-predictive. The data are daily observations from Our World in Data \cite{Mathieu2020mortality,Mathieu2020cases} covering a period from September 1st, 2020, to August 24, 2025. (a) shows the joint phase space of \(\mu\) and \(\beta\). (b) shows \(\mu_{n+1}\) versus \(\mu_n\), and (c) shows \(\beta_{n+1}\) versus \(\beta_n\). Relative to the USA and Europe, the Asian series exhibits a modest increase in dispersion in the one-step relationships, particularly at higher values of transmission. This pattern may reflect the influence of additional causal drivers in this region.}

\FloatBarrier

Across regions, the virulence series exhibits consistently low correlation dimension. Estimates of $D_2(\mu_n)$ are close to unity, and the increase in $D_2([\mu_n,\mu_{n+1}])$ is modest, which indicates strong short-term predictability and a low-dimensional geometric structure. This behavior is visually reflected in the tight one-dimensional manifolds observed in the corresponding scatter plots. In contrast, the transmission series display more variable $D_2$ values (across regions), which suggests more complex underlying dynamics. Geometric information flow further clarifies the asymmetry between virulence and transmission. At both global and regional scales, the information flow from transmission to virulence remains weak. The corresponding $GeoC_{\beta\to\mu}$ values are small and show limited regional variation. By contrast, the flow from virulence to transmission is more pronounced at the global level and in Asia, indicating that contemporaneous virulence contains additional predictive information for subsequent transmission in those settings. The USA and Europe exhibit smaller $GeoC_{\mu\to\beta}$ values, consistent with a weaker influence of virulence on transmission.

The regional scatter plots are consistent with these quantitative findings. The USA and Europe show compact one-step relationships for virulence and comparatively limited transmission dispersion. The Asian series shows dispersion in the one-step transmission relationships, consistent with the larger estimated $GeoC_{\mu\to\beta}$ for this region. These regional patterns should be interpreted cautiously, given differences in epidemic timing, reporting practices, and the potential presence of unobserved co-variates.

\section{Supplementary notes and figures on the LETR study}
\subsection{Study limitations} 
Despite its potential, the LETR framework has several limitations and opportunities for future development. 
Robust Granger-based inference requires sufficiently long and approximately stationary time series, so methodological advances for short or irregular datasets and techniques to mitigate unobserved confounding are required. Accurate causal modeling depends on distinguishing direct from indirect drivers (as illustrated in myxomatosis example), which motivates integrating tools for detecting mediated influences such as causation entropy \cite{sun2014causation}. Extending LETR to include genetic drivers will demand adaptation to high-dimensional genomic data and related extensions of Granger causality \cite{guo2010granger}. Natural systems may be stochastic due to demographic and environmental noise; incorporating stochastic transfer operators and relevant causality metrics is essential \cite{koutsoyiannis2022revisiting}. And LETR may be of use in controlled laboratory settings or in experimental evolution, where the method can more carefully hone in on causal relationships in evolving populations. 

\subsection{Supplementary Figures}
\subsubsection{Causal model for Granger style predictor testing}
\begin{figure}[ht]
    \centering
    \includegraphics[width=0.5\textwidth]{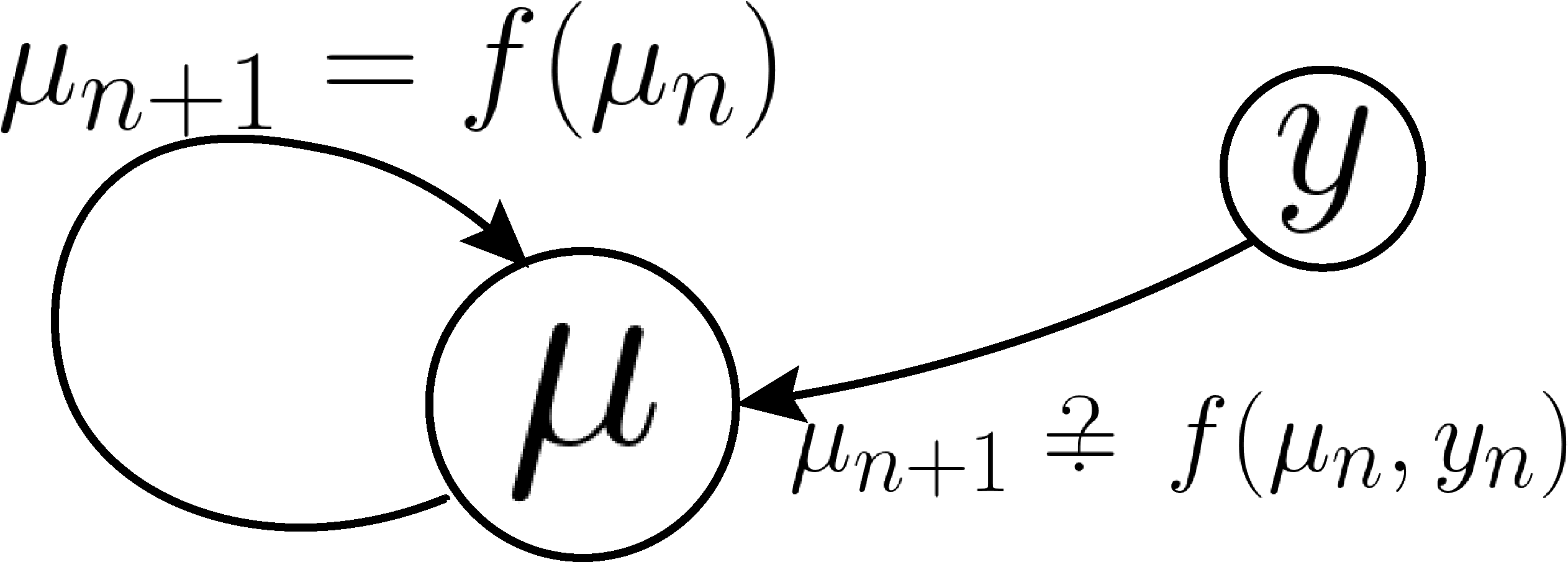}
    \caption{Diagram of the causal model used to test Granger style predictors of a focal trait. The focal trait is denoted by \(\mu\) and observations are indexed by generation \(n\). The baseline hypothesis is that the next-generation value \(\mu_{n+1}\) depends only on the previous value \(\mu_n\), written \(\mu_{n+1}=f(\mu_n)\). The alternative hypothesis is that a measured variable \(y\) at the previous generation improves one step forecasts of the focal trait, written \(\mu_{n+1} =\hspace{-0.08in} ? \hspace{0.08in} f(\mu_n, y_n)\). In practice we test whether including \(y_n\) reduces out of sample forecast error and increases predictive information for \(\mu\). Example candidate variables \(y\) are transmission rate, host immune status, population density, treatment exposure, and environmental co-variates. A positive result should be interpreted as evidence that \(y\) carries predictive information that may reflect a causal influence under the Granger philosophy.}

    \label{fig:CasModel}
\end{figure}

The supplementary \cref{fig:CasModel} presents a schematic of the nested model comparison employed in this study. The null model predicts the focal trait solely from its own lagged value, \(\mu_{n+1}=f(\mu_n)\). The alternative model augments this by conditioning on a candidate predictor \(y_n\) and evaluates whether its inclusion improves one-step forecasts of \(\mu\). In practice, this test is implemented using information gain measures that quantify the predictive improvement attributable to \(y_n\). Illustrative biological examples of such candidate predictors include transmission rate, host immune status, population density, treatment exposure, and environmental co-variates. It is important to emphasize that a positive Granger style result indicates predictive utility rather than definitive mechanistic causation. Inference may be influenced by confounding factors, unobserved drivers, measurement error, or limited temporal resolution. 

\subsubsection{Pathogen passage schematics and ensemble distributions}

\begin{figure}[ht]
    \centering
\begin{tikzpicture}

\node[draw, circle] (pathogen_n) at (0,0) {\(\mu_n, \beta_n\)};
\node[above] at (pathogen_n.north) {$n^{\text{th}}$ generation};

\node[draw, rectangle, minimum width=3cm, minimum height=2cm] (host) at (5,0) {Host};

\node[draw, circle] (pathogen_np1) at (10,0) {\(\mu_{n+1}, \beta_{n+1}\)};
\node[above] at (pathogen_np1.north) {$(n+1)^{\text{th}}$ generation};

\draw[->] (pathogen_n) -- (host.west) node[midway, above] {Enter};
\draw[->] (host.east) -- (pathogen_np1) node[midway, above] {Exit};

\end{tikzpicture}
\caption{Schematic showing a single \(n^{\text{th}}\) generation pathogen entering a host and the resulting \((n+1)^{\text{th}}\) generation pathogen exiting the host. The pathogen in generation \(n\) is characterized by virulence \(\mu_n\) and transmission rate \(\beta_n\). After passing through the host the pathogen in generation \((n+1)\) is characterized by updated values \(\mu_{n+1}\) and \(\beta_{n+1}\) due to mutation, immune response, transmission bottlenecks or other within-host processes. The figure emphasizes that changes between consecutive observations may reflect both evolutionary change during host passage and host-specific effects. In time series that lack a clear generational label one unit of time can be interpreted as a generation with index \(n\).}

    \label{fig:pathogen_cycle}
\end{figure}
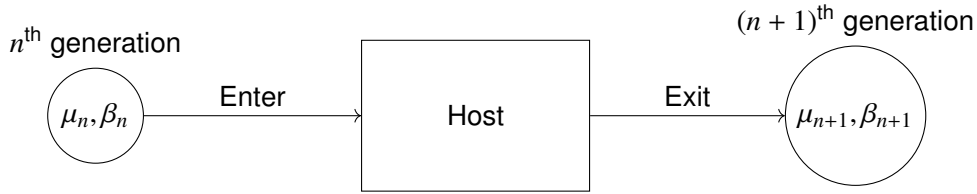

\begin{figure}[h!]
    \centering
    \begin{tikzpicture}[
        % Define styles for nodes
        pathogen/.style={circle, draw, inner sep=2pt},
        host/.style={rectangle, draw, inner sep=4pt},
        normaldist/.style={domain=-3:3, samples=50, smooth, variable=\x},
        logisticdist/.style={domain=-3:3, samples=100, variable=\x}
    ]

    \begin{axis}[
        hide axis,
        width=0.4\textwidth,
        at={(0.1\textwidth,-0.1\textwidth)},
        anchor=north,
    ]
    \addplot[normaldist] {exp(-0.5/0.01*\x*\x) / (sqrt(2*pi))};
    \end{axis}

    \begin{axis}[
        hide axis,
        width=0.4\textwidth,
        at={(0.65\textwidth,-0.1\textwidth)},
        anchor=north,
    ]
    \addplot[logisticdist] {exp(-0.5*\x*\x) / (sqrt(2*pi)))};
    \end{axis}
    
    \foreach \y/\m in {0/1, -2/2, -4/3,-8/m} {
        \node[pathogen] (pathogen) at (3,\y) {};
        \node[right] at (2.5,\y-0.5) {\(\mu_n^{\m} \), \(\beta_n^{\m} \)};
        \node[host] (host) at (6,\y) {};
        \node[right] at (5.5,\y-0.5) {$\text{Host}_\m$};
        \draw[->] (pathogen) -- (host);
        \node[pathogen] (pathogen_exit) at (9,\y) {};
        \node[right] at (8.5,\y-0.5) {\(\mu_{n+1}^{\m}\), \(\beta_{n+1}^{\m}\)};
        % Arrow from host to pathogen_exit
        \draw[->] (host) -- (pathogen_exit);
    }
    \node[right] at (5.5,-6) {\vdots};
    % Labels
    \node[below] at (0.1\textwidth,-0.35\textwidth) {Initial Distribution (eg:\(\mu_n\))};
    \node[below] at (0.65\textwidth,-0.35\textwidth) {Ending Distribution (eg:\(\mu_{n+1}\))};
    
    \end{tikzpicture}
\caption{Schematic representation of pathogen passage across hosts. 
Pathogens from generation \(n\) enter hosts carrying trait values (for example, transmission rate \(\beta\) or virulence \(\mu\)). 
After undergoing within-host processes such as mutation and ecological interactions, they exit the hosts as pathogens of generation \(n+1\) with potentially altered traits. 
Illustrated distributions show how the values of a trait may shift from entry (generation \(n\)) to exit (generation \(n+1\)).}

    \label{fig:pathogen_multiple_hosts}
\end{figure}
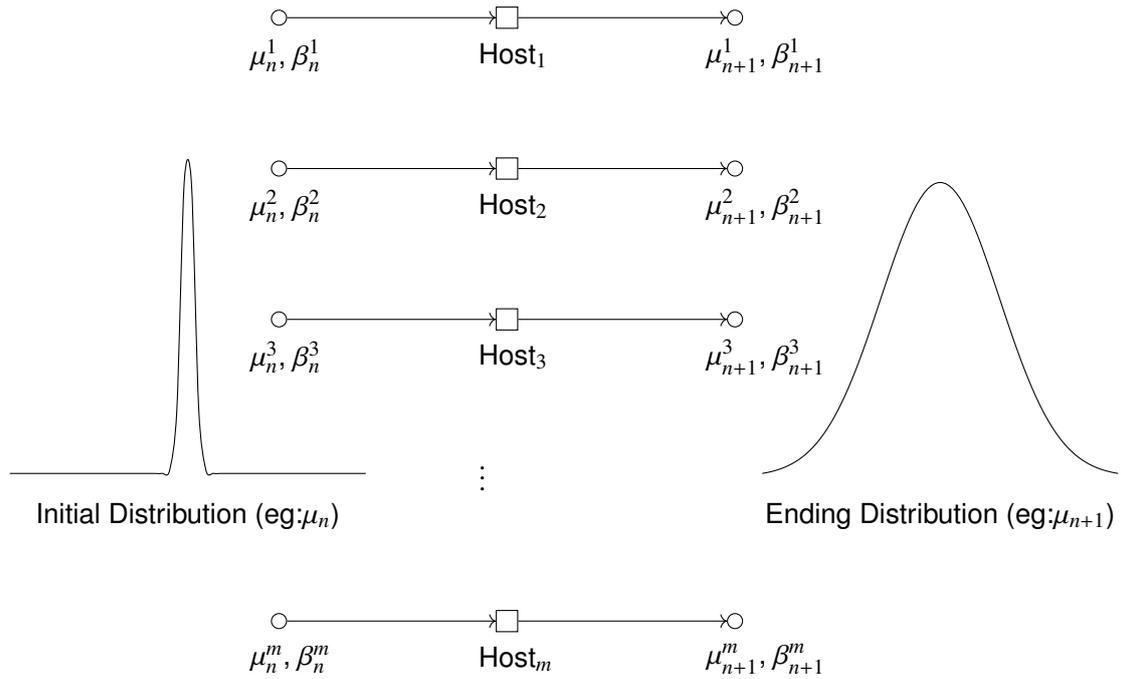

\cref{fig:pathogen_cycle} depicts a single infection event in which a pathogen of generation \(n\) with traits \(\mu_n\) and \(\beta_n\) enters a host and exits as generation \(n+1\) with updated traits \(\mu_{n+1}\) and \(\beta_{n+1}\). This diagram highlights that observed changes between successive records may arise during host passage through mutation, selection, transmission bottlenecks and host specific ecological effects. \cref{fig:pathogen_multiple_hosts} adopts an ensemble perspective and shows how repeated passages across many hosts reshape the population level distribution of a focal trait. Together, these separate figures clarify the link between within host processes and population level evolution and motivate the focus on one step transition densities and invariant densities under a discrete update map. For empirical time series that lack an explicit generational label, one unit of observation may be interpreted as a generation indexed by \(n\).

\subsubsection{Simulated time series data synthetic example and causal model}\label{asec:timeSdata}

\begin{figure}[ht!]
    \centering
    \includegraphics[width=0.5\linewidth]{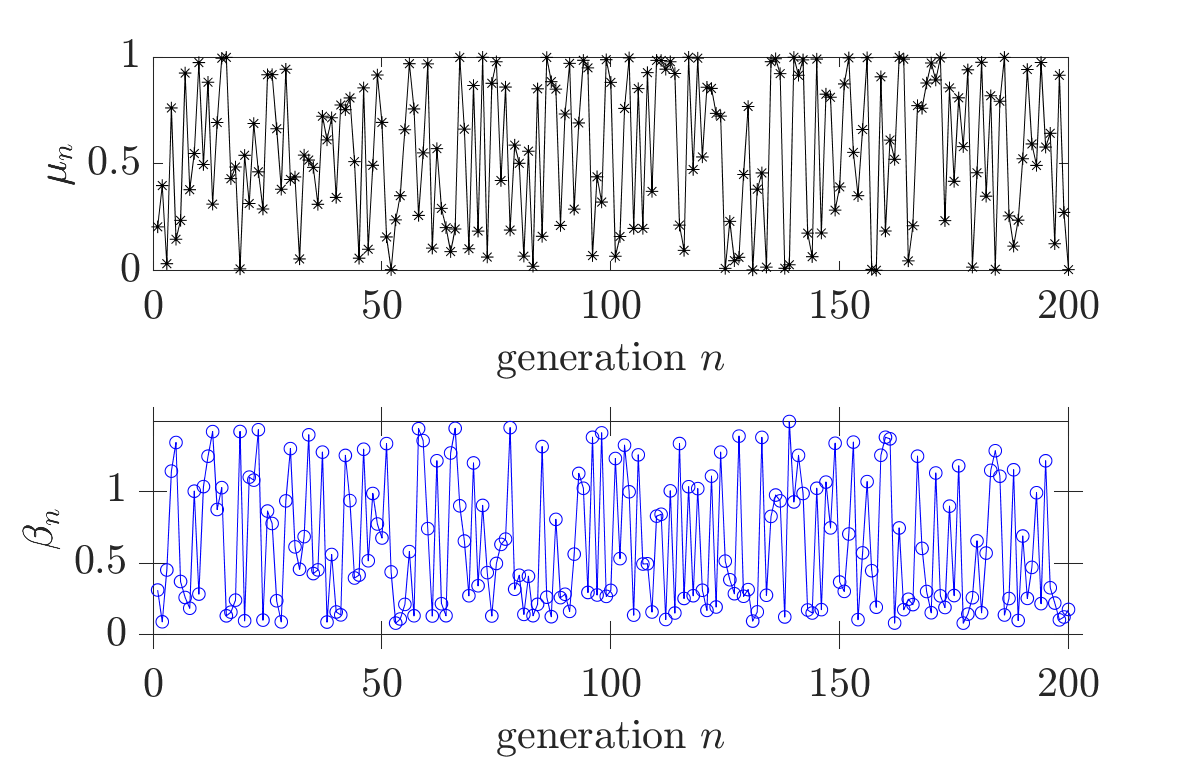}
\caption{Simulated time series of virulence \(\mu_n\) and transmission \(\beta_n\) across 200 generations. The trajectories appear noisy but were generated by deterministic update rules chosen to mimic mechanistic links between virulence and transmission, following the spirit of Dwyer et al. \cite{dwyer1990simulation}. The generating map used in the simulation is
$
\mu_{n+1} = a\,\mu_n(1-\mu_n),\qquad
\beta_{n+1} = 1.5\,\operatorname{sech}^2(\mu_n+\beta_n)$. The \(\mu\) update is the logistic map in its chaotic regime ($a\approx 4$) and can produce complex, irregular trajectories even in the absence of external noise. The \(\beta\) update couples transmission to both the current virulence level and the current transmission level through a nonlinear bell-shaped bounded function, producing dependent fluctuations in transmission. These rules reproduce features seen in empirical pathogen data such as irregular trajectories, non-Gaussian trait distributions, and temporal predictability of virulence that can influence transmission.}

    \label{fig:munBetan}
\end{figure}

\cref{fig:munBetan} display the simulated time series used throughout this study. The generating maps employed in these simulations are
\[
\mu_{n+1} = a\,\mu_n(1-\mu_n), \qquad
\beta_{n+1} = 1.5\,\operatorname{sech}^2(\mu_n+\beta_n),
\]
with the parameter \(a\) chosen near four so that the \(\mu\) update operates in the chaotic regime originally popularized in ecological applications by May \cite{may1976simple}. The \(\beta\) update implements a bounded response function of \(\mu_n\) and \(\beta_n\) so that transmission remains within a biologically plausible range. These deterministic rules were chosen to produce irregular looking trajectories for pedagogical purposes while preserving interpretable links between virulence and transmission.

\begin{figure}[ht]
    \centering
    \includegraphics[width=0.5\textwidth]{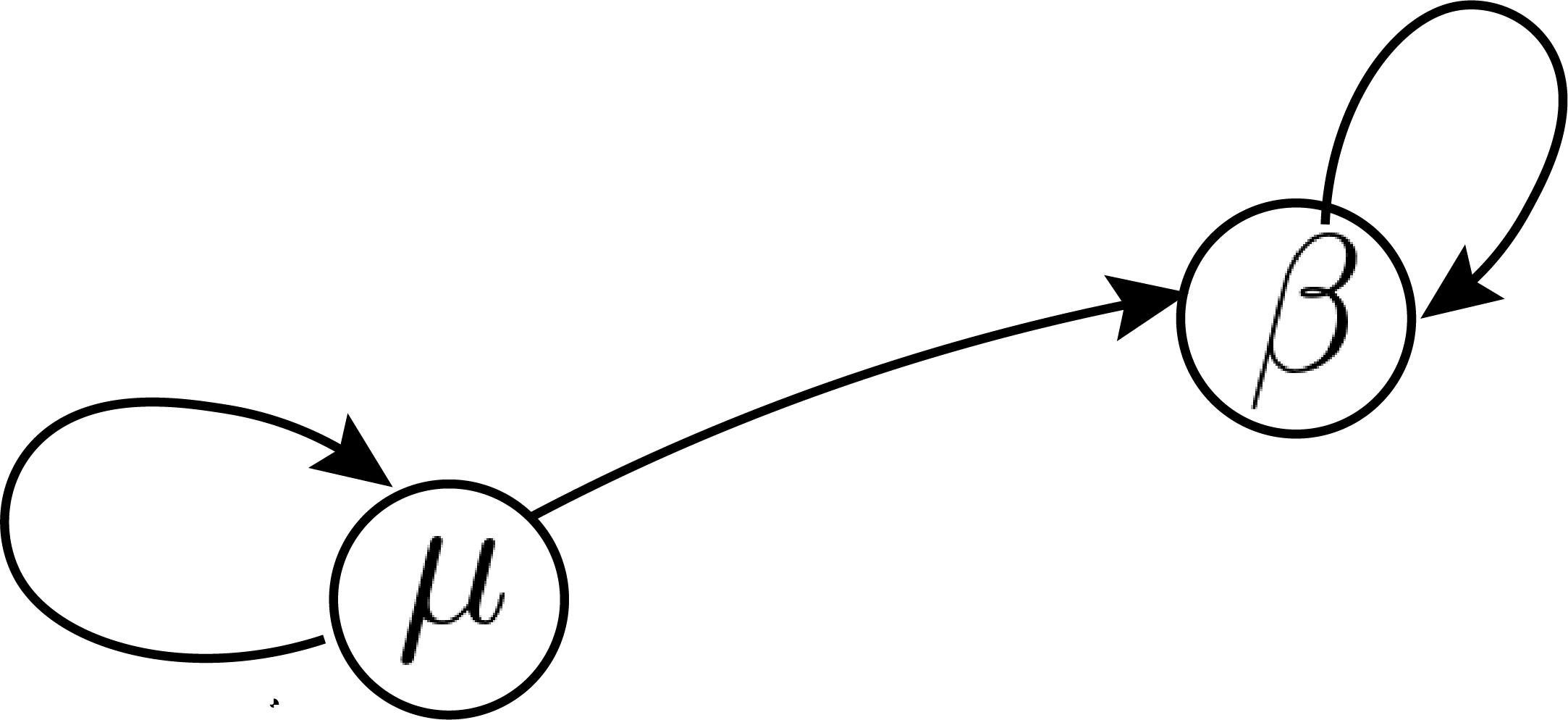}
    \caption{Causal model summarizing the inferred relationships from the simulated dataset and supporting empirical evidence. 
The analysis reveals an asymmetric relationship in which past virulence (\(\mu_n\)) substantially improves the prediction of future transmission (\(\beta_{n+1}\)), whereas the reverse effect is weak. 
This conclusion is quantitatively supported by the geometric causal inference measures reported in the main text (e.g., \(GeoC_{\mu \to \beta} \approx 0.46\) and \(GeoC_{\beta \to \mu} \approx 0.02\) in these simulations). 
The dominant directional link from virulence to transmission reflects the mechanistic assumptions embedded in the simulation. 
The generating map employed in this simulation was 
\(
\mu_{n+1} = 4\,\mu_n(1-\mu_n), \qquad \beta_{n+1} = 1.5\,\operatorname{sech}^2(\mu_n + \beta_n),
\)
which capture the deterministic update rules underlying the observed dynamics. 
This framework can be extended to experimental or observational data to investigate causal relationships in real biological systems.}

    \label{fig:CasModelMuBeta}
\end{figure}

In \cref{fig:CasModelMuBeta}, we present the inferred causal model derived from the simulated dataset. The figure illustrates the asymmetric relationship between virulence (\(\mu_n\)) and transmission (\(\beta_{n+1}\)), showing that past virulence substantially improves the prediction of future transmission, whereas the reverse effect is negligible. These relationships are quantitatively supported by the geometric causal measures reported in the main text, with \(GeoC_{\mu \to \beta} \approx 0.67\) and \(GeoC_{\beta \to \mu} \approx 2 \times 10^{-5}\). The dominant directional link from virulence to transmission reflects the deterministic generating maps employed in the simulation, specifically \(\beta_{n+1} = 1.5\,\operatorname{sech}^2(\mu_n + \beta_n)\). This model illustrates how LETR captures mechanistic assumptions embedded in simulated data and provides a robust framework for investigating causal relationships in experimental or observational biological systems.

\section{Dynamics of virulence as a logistic map}\label{aSec:LenVir}

In epidemiology, the case mortality rate is often used as a proxy for virulence \cite{dwyer1990simulation}. Analogous to how the number of infected cases over time can be modeled using a logistic map \cite{sugihara1990nonlinear}, the case mortality rate can likewise be conceptualized as evolving according to a logistic map. This idea can be further supported through data-driven analysis. Here, we use historical real-world data on \emph{poliomyelitis} to demonstrate that its virulence dynamics follow a logistic map. Furthermore, after investigating the poliomyelitis data we did not identify any additional causal variables. Hence we now focus exclusively on virulence as the pathogen's trait of interest, assuming no influence from transmission or other parameters on its dynamics. We further assume that virulence evolves across generations according to the logistic map with hyperparameter \(0 < a \le 4\).

 \cref{eq:lgMapEx} illustrates the logistic map relationship between the \(n^{\text{th}}\) and \((n+1)^{\text{th}}\) generations of the pathogen's virulence value.
\begin{align}\label{eq:lgMapEx}
    \mu_{n+1}=a \mu_n (1-\mu_n).
\end{align}
 Depending on the parameter value \(a\), the map can demonstrate a range of behaviors, from stable fixed points to periodic orbits and chaotic dynamics. It is particularly notable for undergoing period doubling bifurcations as \(a\) increases. The chaotic behavior of the map, especially when \(a = 4\) \cite{devaney1989introduction}, is of special interest due to its relevance in practical scenarios.

\subsection{Chaos in the logistic map and long-term trait distributions}\label{sec:apndixChaos}

To offer a concise overview of chaotic behavior, we summarize Erik Bollt's (1997) \cite{bollt1997controlling} explaination. People often ask, ``What is chaos good for?'' This question underscores the intriguing nature of chaos \cite{gutzwiller2013chaos, smith1993dynamic, bollt1997controlling}. Although chaos effectively characterizes the complexity of the world around us, its importance reaches beyond mere description. Chaos can serve as a powerful model for understanding our environment and, moreover, can be controlled and harnessed \cite{bollt1997controlling}. The term ``chaos'' is often misunderstood due to the disparity between its common English usage and its mathematical definition. In everyday language, chaos is typically associated with ``extreme disorder or confusion'' \cite{ref1}. In contrast, the mathematical concept of chaos suggests that what appears to be random behavior may arise from an underlying deterministic and structured process. According to the well-accepted definition by Devaney \cite{devaney2018introduction}, a one-dimensional map \(f\) of the form \(x_{n+1} = f(x_n)\) is chaotic if it displays sensitive dependence on initial conditions, exhibits topological transitivity (equivalent to the existence of a dense orbit), and contains an infinite number of periodic orbits \cite{bollt1997controlling, bollt1995targeting}. 

\begin{figure}[ht!]
        	\centering
        	\includegraphics[width=0.8\textwidth]{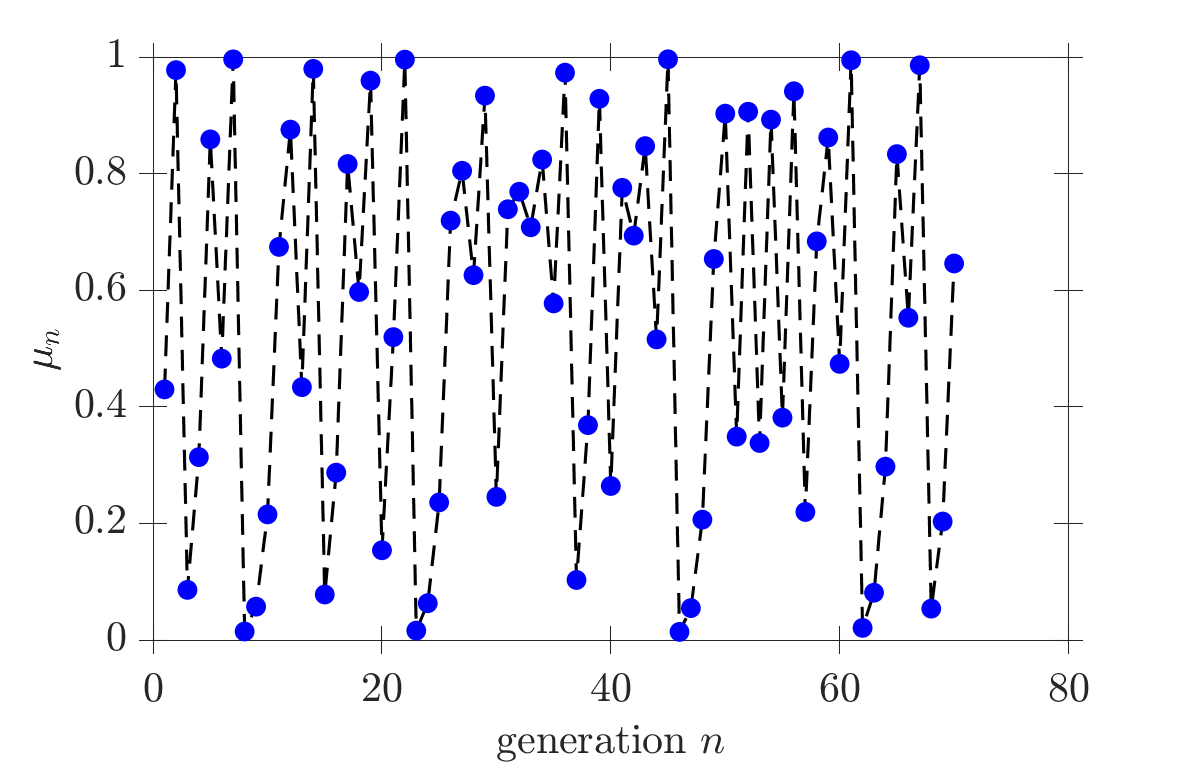}
            \caption{Chaotic variation in a pathogen trait generated by a simple deterministic model. The time series shows the trajectory of a hypothetical focal trait ($\mu$) across successive pathogen generations. Although the pattern appears noisy, it is generated by the logistic map, a discrete nonlinear recurrence relation popularized by the biologist Robert May in a landmark 1976 Nature paper\cite{may1976simple}, demonstrating how simple deterministic rules can produce complex, irregular dynamics. This example illustrates how pathogen traits ($\mu$) may appear to fluctuate stochastically without any external randomness, reflecting deterministic ecological and evolutionary feedbacks.}
        	\label{fig:mutdata}
\end{figure}

The surprising insight of chaos theory is that data that appear to be randomly generated, such as the time series presented in \cref{fig:mutdata}, are not necessarily random \cite{bollt1997controlling}. Therefore, the logistic map with \(a = 4\) is employed in this study to model virulence, demonstrating the systems chaotic behavior and enabling detailed investigation of seemingly random data. Utilizing this chaotic map, we illustrate the evolution of ensembles (densities) of initial conditions under the dynamical system. As discussed in the methods section, the analysis of ensemble evolution can effectively be concluded using the Frobenius Perron operator theory. This theoretical framework is broadly applicable, accommodating various deterministic or random maps beyond the logistic map example. Readers can explore any deterministic or random map for discussions on transfer operator theory or ensemble evolution. 

\begin{figure}[ht!]
    \centering
    \begin{subfigure}{0.45\textwidth}
        \centering
        \includegraphics[width=\textwidth]{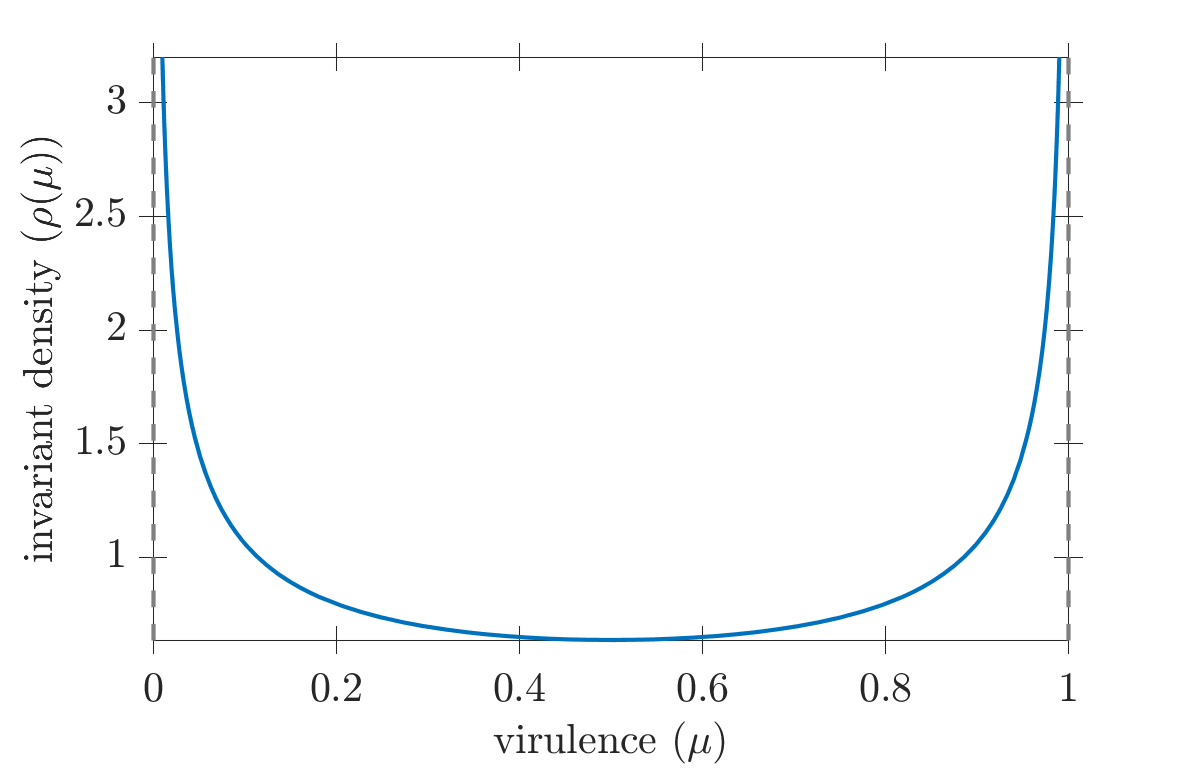}
        \caption{}
        \label{fig:InvDenLogMap}
    \end{subfigure}
    \begin{subfigure}{0.45\textwidth}
        \centering
    \includegraphics[width=\textwidth]{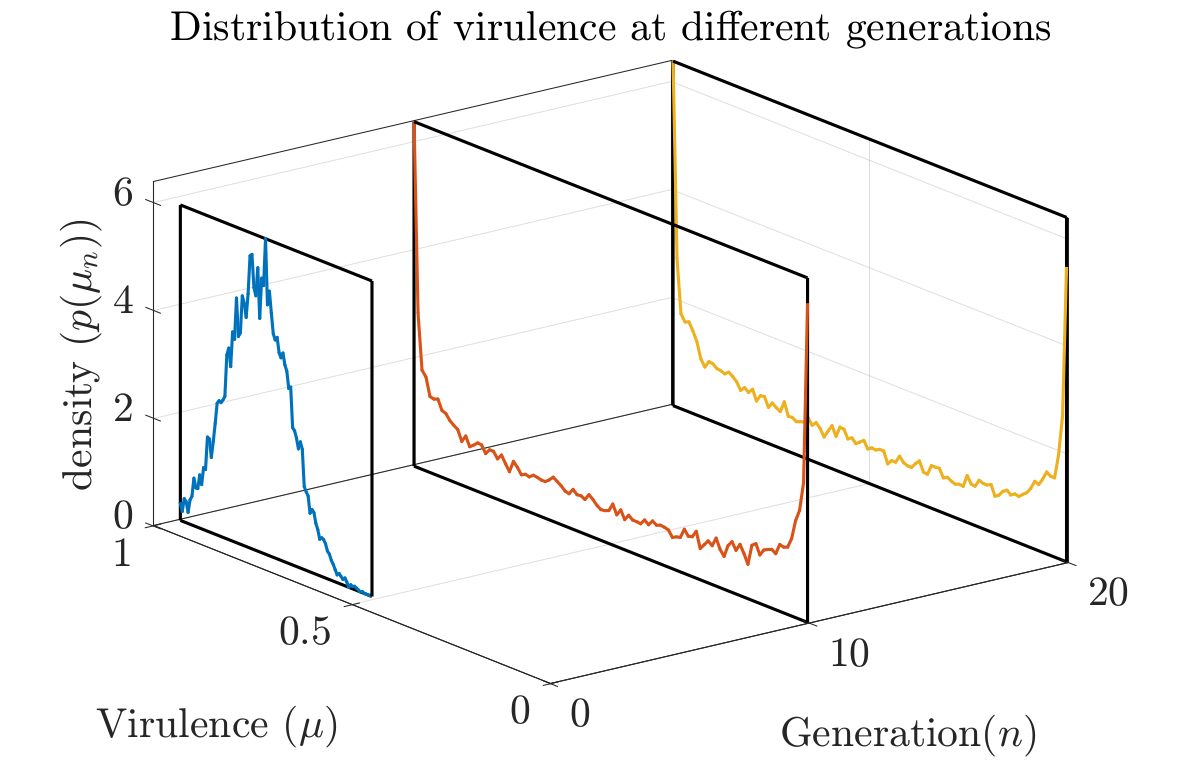}
        \caption{}
        \label{fig:DenMuGen}
    \end{subfigure}
    
\caption{Long term dynamics of pathogen virulence modeled by the logistic map.  
(a) The invariant density of the logistic map with \(a = 4\) is given by \(\rho(\mu) = \frac{1}{\pi\sqrt{\mu(1-\mu)}}\), indicating that trait values near 0 and 1 are more frequently visited after many generations.  
(b) The evolution of the trait distribution over successive generations demonstrates convergence to this invariant density.  
This pattern suggests that, in the context of virulence, pathogen populations are more likely to evolve towards either very low or very high virulence in the long-term, with a significantly lower population density in the intermediate range. This pattern is reminiscent of certain ecological and evolutionary scenarios in host–parasite systems, where selective pressures may favor either very low or very high virulence strategies over intermediate values \cite{baalen1998coevolution, hite2018resource}. This illustrates how simple deterministic models can capture complex evolutionary patterns in pathogen traits.}

    \label{fig:ActOfFP}
\end{figure}

The logistic map with \(a = 4\) is renowned for its unique absolutely continuous invariant density function \cite{lasota1998chaos, bollt2000controlling, surasinghe2024learning} (see \cref{fig:InvDenLogMap}), represented by
\begin{align}\label{eq:InvDenLmap}
    \rho(\mu) = \frac{1}{\pi\sqrt{\mu(1-\mu)}}.
\end{align}
This closed-form expression of the invariant density for the logistic map enables detailed analysis. Regardless of the initial distribution of the variable, the long-term distribution converges to this invariant density (\cref{eq:InvDenLmap}), as shown in \cref{fig:DenMuGen}. As observed in \cref{fig:ActOfFP}, the density is higher around 0 and 1 than at other values. In the context of virulence, this implies that, regardless of the initial distribution of virulence values in the pathogen population, the long-term trend is to see the majority of the population leaning towards either the lowest virulence (zero) or the highest virulence (one), with a significantly lower density in the intermediate range.

\clearpage
{\small
\printbibliography[title={Supporting Information References}]
}
\end{refsection}

% \clearpage
% {\small 
% \bibliography{refs}}

% \setcounter{figure}{0}
% \renewcommand{\thefigure}{B\arabic{figure}}
% \setcounter{table}{0}
% \renewcommand{\thetable}{B\arabic{table}}

\end{document}